\documentclass[]{iopart}
\usepackage{graphicx}
\usepackage{cite}
\usepackage{iopams}

\graphicspath{{.}}

\newcommand{\spterm}[3]{\ensuremath{^{#1}#2_{#3}}}
\newcommand{\chspc}{\hspace{0.1em}}
\newcommand{\dmo}[4]{{#1}$_{#2}${#3}$_{#4}$MnO$_3$}
\newcommand{\clebsch}[6]{{\left(
      \begin{array}{cc|c}
        {#1} & {#3} & {#5} \\
        {#2} & {#4} & {#6}
      \end{array}\right)}}

\begin{document} 

\title[]{Microscopic modelling of doped manganites} 

\author{Alexander Wei{\ss}e\dag{} and Holger Fehske\ddag{}}
\address{\dag\ School of Physics, The University of New South Wales, 
  Sydney NSW 2052, Australia}
\address{\ddag\ Institut f\"ur Physik, Ernst-Moritz-Arndt 
  Universit\"at Greifswald, 17487 Greifswald, Germany}
\ead{aweisse@phys.unsw.edu.au, holger.fehske@physik.uni-greifswald.de} 

\begin{abstract} 
  Colossal magneto-resistance manganites are characterised by a
  complex interplay of charge, spin, orbital and lattice degrees of
  freedom.  Formulating microscopic models for these compounds aims at
  meeting to conflicting objectives: sufficient simplification without
  excessive restrictions on the phase space. We give a detailed
  introduction to the electronic structure of manganites and derive a
  microscopic model for their low energy physics. Focussing on short
  range electron-lattice and spin-orbital correlations we supplement
  the modelling with numerical simulations.
\end{abstract} 
\submitto{\NJP} 


\section{Introduction}

Mixed-valence manganese oxides of perovskite structure have been
studied for more than fifty years~\cite{JS50,WK55}. However, many of
their peculiar properties are still rather poorly understood. About a
decade ago the observation of the so-called colossal magneto-resistance
effect (CMR)~\cite{KSKMH89,HWHSS93,JTMFRC94} moved these materials
again into the focus of intense research
activity~\cite{CVM99,TT99,DHM01}. It turned out soon that the complex
electronic and magnetic properties of the manganites as well as their
rich phase diagram depend on a close interplay of almost all degrees
of freedom known in solid state physics, namely itinerant charges,
localised spins, electronic orbitals, and lattice vibrations. To
separate important and less important degrees of freedom and to work
out the essential low energy physics are therefore two particularly
crucial factors for any microscopic description of these interesting
compounds. In the present article we give a detailed introduction to
the electronic structure of manganites and derive a microscopic model,
which includes the dynamics of charge, spin, orbital and lattice
degrees of freedom on a quantum mechanical level. We also discuss
potential further simplifications and complement the work with a
review of previous numerical studies of the derived model.

\section{Crystal structure and symmetries}
\begin{figure}
  \begin{center}
    \includegraphics[height=0.25\textheight]{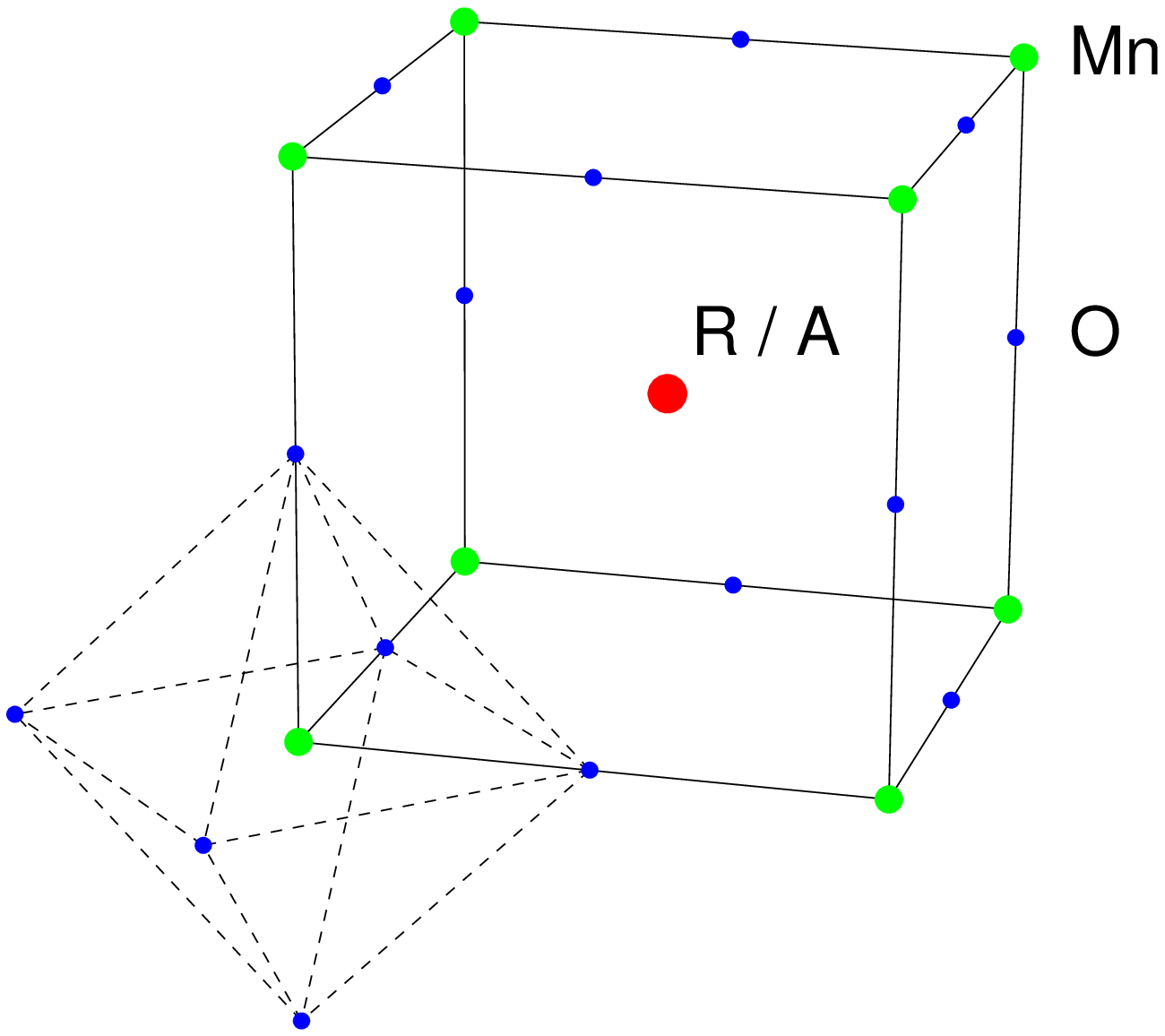}
    \hspace{0.1\linewidth}
    \includegraphics[height=0.25\textheight]{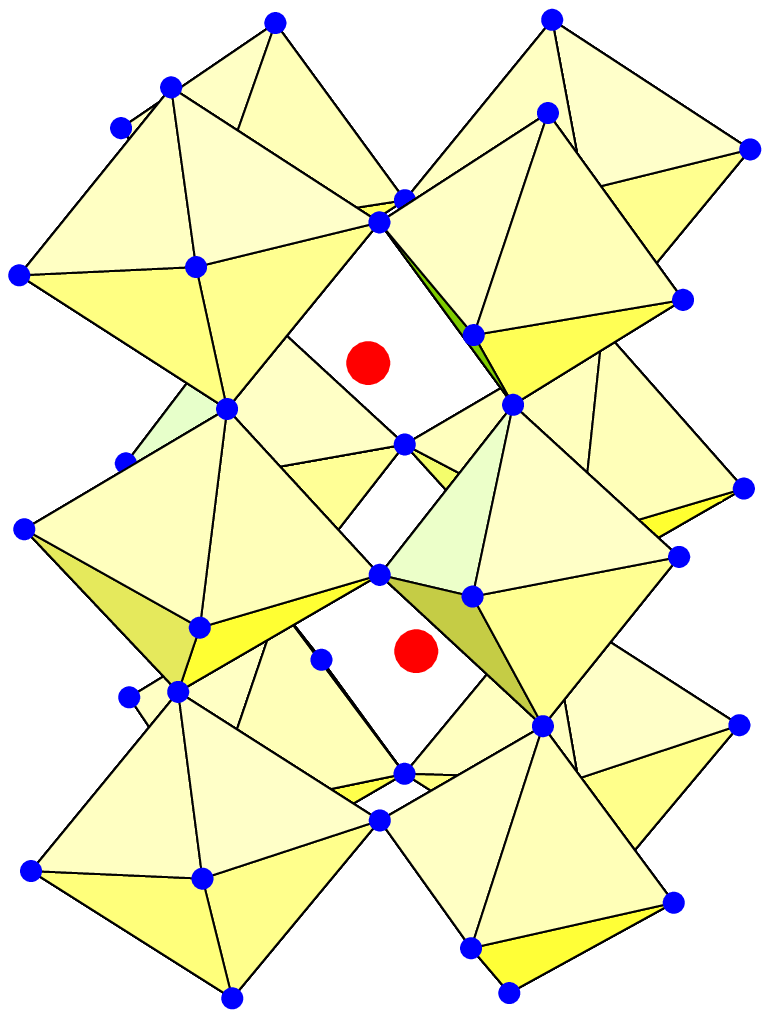}
  \end{center}
  \caption{Left: Idealised perovskite structure (space group $Pm3m$). 
    Right: Real structure of \dmo{La}{}{}{} at $T=1.5\,K$
    (orthorhombic, space group $Pnma$)~\cite{Moea96}.}\label{figperow}
\end{figure}
The structural phase diagrams observed for CMR manganites are almost
as rich as their electronic and magnetic counterparts (for a review
see, e.g., Ref.~\cite{CVM99}). Consider, for example, the undoped
parent compound \dmo{La}{}{}{}. At low temperature its structure is
characterised by the orthorhombic space group $Pnma$~\cite{Moea96},
but doping with strontium, \dmo{La}{1-x}{Sr}{x}, will transform the
system to the triclinic space group $R\bar{3}c$ at $x\approx
0.175$~\cite{Urea95,Moea96}. On the other hand, similar structural
transitions occur with increasing temperature~\cite{RCea98}.  However,
nearly all of these crystal arrangements can be understood in terms of
distortions of the ideal perovskite structure shown in
\Fref{figperow}. Here each manganese ion is surrounded by an
octahedron of oxygen atoms, whereas the large rare earth (R) or
alkaline earth atoms (A), used as dopants, occupy the centre of the
cube formed by the manganese sites.

\begin{table} 
  \caption{\label{tabohgen}The four symmetry operations that generate 
    the cubic point group $O_h$, the corresponding notation as well 
    as the action on the position vector.} 
  \begin{indented} 
  \item[]\begin{tabular}{llrcl} 
      \br
      Operation & Symbol & \multicolumn{3}{l}{Coordinate transformation}\\
      \mr
      inversion & $I$ &
      $(x,y,z)$ & $\rightarrow$ & $(-x,-y,-z)$\\[1mm]
      4-fold rotation about $x$-axis & $C_4^x$ &
      $(x,y,z)$ & $\rightarrow$ & $(x,-z,y)$\\[1mm]
      4-fold rotation about $z$-axis & $C_4^z$ &
      $(x,y,z)$ & $\rightarrow$ & $(-y,x,z)$\\[1mm]
      3-fold rotation about diagonal & $C_3^d$ &
      $(x,y,z)$ & $\rightarrow$ & $(z,x,y)$\\
      \br
    \end{tabular} 
  \end{indented} 
\end{table}

For a theoretical description it is thus quite natural to start from
the above ideal cubic structure, and to account for deviations by
including (dynamical) lattice distortions, i.e., phonons, into the
microscopic model. Since the cubic site symmetry of the manganese is
particularly important, and since its properties and the corresponding
notation are frequently used throughout the article, let us first
recall some of the basic features of the cubic point symmetry group
$O_h$. The group $O_h$ consists of 48 symmetry operations which can
all be generated by the inversion of space and the 3 basic rotations
listed in \Tref{tabohgen}. Since the rotations commute with the
inversion the group $O_h$ can be represented as a direct product, $O_h
= O\otimes C_i$, of the inversion group $C_i$ and the group $O$, which
is formed by the rotations only. With respect to inversion every
function can be decomposed into an even ($g$) and an odd ($u$) part,
which reflects the two irreducible representations of $C_i$ with
characters $1$ and $-1$, respectively. Similarly, every irreducible
representation of $O$ leads to an even and an odd irreducible
representation of $O_h$. Given the five irreducible representations of
$O$ listed in \Tref{tabirrep} we thus find ten irreducible
representations of the full cubic point group $O_h$, denoted by
$A_{1g}$, $A_{1u}$, $A_{2g}$, etc. Here we follow the notation
commonly used in the literature (see, e.g., the textbook of
Griffith~\cite{Gr71}).

\begin{table} 
  \caption{\label{tabirrep}The five irreducible representations of 
    the cubic group $O$, the notation used for standard basis elements, 
    examples of corresponding functions, and transformation properties 
    with respect to the group generators. Note that we rotate the functions, 
    not the coordinate system.}
  \begin{indented} 
  \item[]\begin{tabular}{ccccccc} 
      \br
      irred. & & basis  &  & \multicolumn{3}{l}{transformation properties}\\
      repr. & dim. &  element & example & $C_4^z$ & $C_4^x$ & $C_3^d$\\
      \mr
      $A_1$ & 1 & $a_1$ & $1$ & $a_1$ & $a_1$ & $a_1$ \\[2mm]
      $A_2$ & 1 & $a_2$ & $xyz$ & $-a_2$ & $-a_2$ & $a_2$ \\[2mm]
      $E$ & 2 & $\theta$ & $3z^2 - r^2$ & $\theta$ &
      $-\frac{1}{2}\theta-\frac{\sqrt{3}}{2}\varepsilon$ &
      $-\frac{1}{2}\theta+\frac{\sqrt{3}}{2}\varepsilon$ \\
      & & $\varepsilon$ & $x^2-y^2$ & $-\varepsilon$ &
      $-\frac{\sqrt{3}}{2}\theta+\frac{1}{2}\varepsilon$ &
      $-\frac{\sqrt{3}}{2}\theta-\frac{1}{2}\varepsilon$ \\[2mm]
      $T_1$ & 3 & $x$ & $x$ & $y$ & $x$ & $y$\\
      & & $y$ & $y$ & $-x$ & $z$ & $z$\\
      & & $z$ & $z$ & $z$ & $-y$ & $x$\\[2mm]
      $T_2$ & 3 & $\xi$ & $yz$ & $-\eta$ & $-\xi$ & $\eta$ \\
      & & $\eta$ & $zx$ & $\xi$ & $-\zeta$ & $\zeta$ \\
      & & $\zeta$ & $xy$ & $-\zeta$ & $\eta$ & $\xi$\\
      \br
    \end{tabular}
  \end{indented} 
\end{table}

As we know, for example from the coupling of angular momenta, the
product of elements belonging to two irreducible representations in
general leads to a function that belongs to a reducible representation,
which can again be decomposed into irreducible representations. For
spins the corresponding coupling coefficients are known as the Clebsch
Gordan coefficients, and, of course, for the cubic group analogous
coefficients were calculated and are listed, e.g., in
Ref.~\cite{Gr71}. For illustration consider the product $E\otimes
T_1$, which leads to a six-dimensional reducible representation that
can be decomposed into $T_1$ and $T_2$. In terms of the above basis
functions we find
\begin{eqnarray}
  x' = -\frac{1}{2}\theta\otimes x 
  + \frac{\sqrt{3}}{2} \varepsilon\otimes x\,, \qquad\qquad & 
  \xi' =  -\frac{\sqrt{3}}{2}\theta\otimes x
  - \frac{1}{2} \varepsilon\otimes x\nonumber{}\,,\\
  y' = -\frac{1}{2}\theta\otimes y
  - \frac{\sqrt{3}}{2} \varepsilon\otimes y\,, \qquad\qquad & 
  \eta' = \frac{\sqrt{3}}{2}\theta\otimes y
  - \frac{1}{2} \varepsilon\otimes y\,,\\
  z' = \theta\otimes z\,, \qquad\qquad &
  \zeta' = \varepsilon\otimes z\,,\nonumber{}
\end{eqnarray}
and obviously the new (dashed) functions fulfil the transformation
properties of basis functions of $T_1$ and $T_2$.

\section{Coulomb interaction and local electronic structure}\label{sec_ion}
Considering the end members of the manganite series \dmo{R}{1-x}{A}{x}
at doping $x=0$ and $1$, respectively, within a simplified ionic
description the different constituents are assigned the formal
valences R$^{3+}$\chspc{}Mn$^{3+}$\chspc{}O$_{3}^{2-}$ and
A$^{2+}$\chspc{}Mn$^{4+}$\chspc{}O$_{3}^{2-}$. This corresponds to
always completely filled oxygen $p$-bands and, with increasing $x$, to
hole doping of the manganese $d$-shell, $d^4\rightarrow d^3$. To some
degree already the early doping dependent measurements of the local
magnetic moment in \dmo{La}{1-x}{Ca}{x} by Wollan and
Koehler~\cite{WK55} confirm this general picture, but also recent band
structure calculations~\cite{PS96a,SPV96} and spectroscopic
experiments~\cite{Abea92,CMS93,Saea95b} are indicative of manganese
$3d$ conduction bands. Although these bands are partially subject to
hybridisation with neighbouring oxygen $2p$ states, it is thus
reasonable and common to consider Mn $d$ electrons as the starting
point of a microscopic modelling, and proceed along the line which is
called the weak-field coupling scheme~\cite{Gr71}.

\begin{figure}
  \begin{minipage}{0.6\linewidth}
    \begin{center}
      \includegraphics[width = 0.8\linewidth]{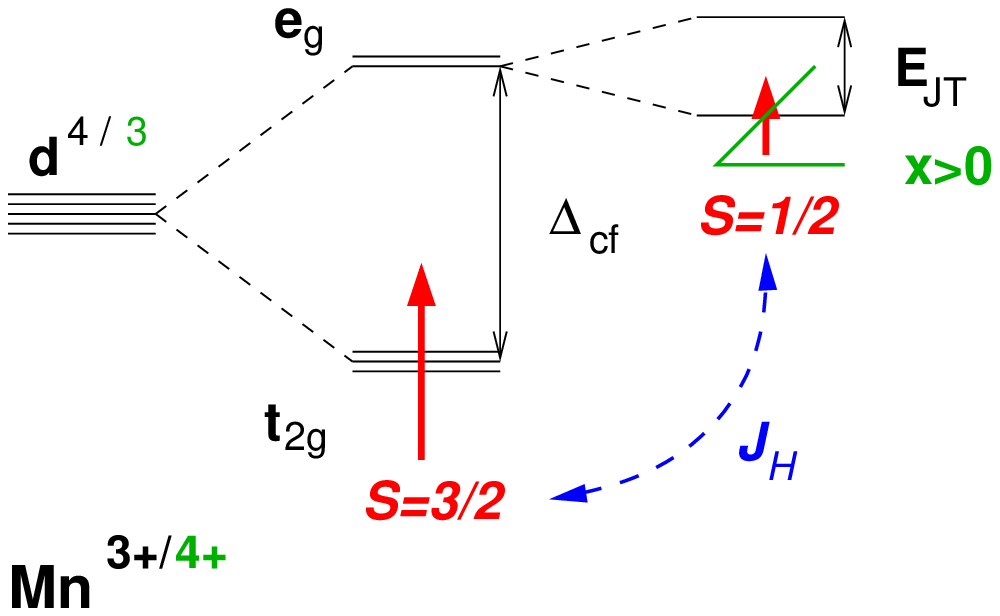}
    \end{center}
  \end{minipage}
  \begin{minipage}{0.4\linewidth}
    \begin{center}
      \includegraphics[width = 0.8\linewidth]{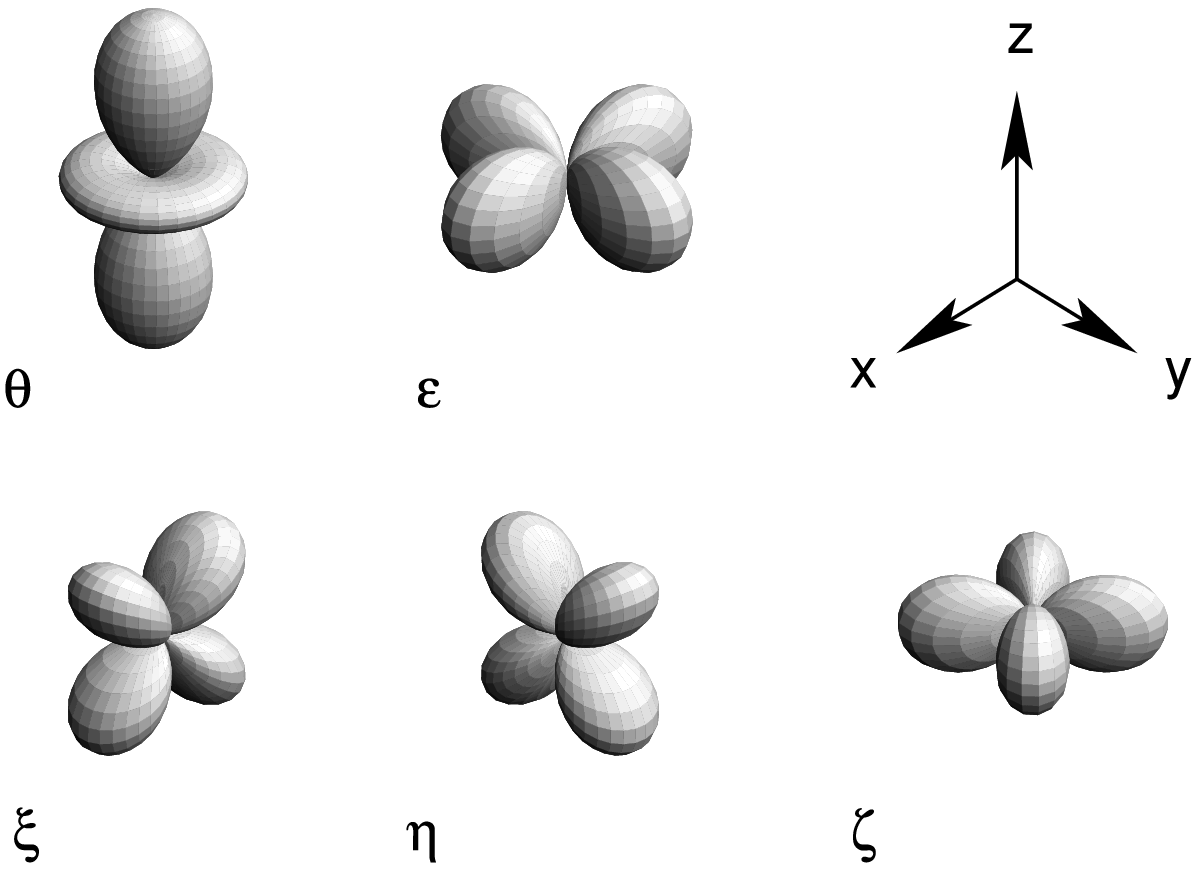}
    \end{center}
  \end{minipage}
  \caption{Left: The local electronic structure of Mn $3d$ electrons in a 
    cubic crystal field. Right: Angular structure of the $e_g$ and
    $t_{2g}$ single electron wave functions.}
  \label{figlevel}
\end{figure}

The spherical harmonics $Y_{l,m}$ with $l=2$, which describe the
angular part of the $d$ shell, are all even with respect to
inversions. Comparing with \Tref{tabirrep} we thus find that these
functions can be combined to yield basis functions of the irreducible
representations $E_g$ and $T_{2g}$ of $O_h$,
\begin{eqnarray}
\fl    \theta  = Y_{2,0}
     = \sqrt{\frac{5}{16\pi}}\frac{3z^2-r^2}{r^2}\,, \qquad &
    \xi  =  i\, \frac{Y_{2,+1} + Y_{2,-1}}{\sqrt{2}} 
     =  \sqrt{\frac{15}{4\pi}}\frac{yz}{r^2}\,,\nonumber{}\\
\fl     \varepsilon  =  \frac{Y_{2,+2} + Y_{2,-2}}{\sqrt{2}} 
    = \sqrt{\frac{5}{16\pi}} \frac{x^2-y^2}{r^2}\,, \qquad &
    \eta   =  -\,\frac{Y_{2,+1} - Y_{2,-1}}{\sqrt{2}} 
    =  \sqrt{\frac{15}{4\pi}}\frac{zx}{r^2}\,,\\
\fl     & 
    \zeta  =  -i\,\frac{Y_{2,+2} - Y_{2,-2}}{\sqrt{2}} 
    =  \sqrt{\frac{15}{4\pi}}\frac{xy}{r^2}\,.\nonumber{}
\end{eqnarray}
Consequently, within the cubic crystal field the five degenerate $d$
levels split into an $E_g$ orbital doublet and a $T_{2g}$ orbital
triplet. Taking into account the spatial structure of the above
wavefunctions and the overlap with neighbouring oxygen electrons, we
find that the $t_{2g}$ orbitals are lower in energy compared to
$e_{g}$. The magnitude of this crystal field splitting, $\Delta_{\rm
  cf}$, varies with the occupancy of the levels, typical values for
the manganites are of the order of 1.5 to 2.5~eV~\cite{Abea92}.

So far we considered only the single-electron picture, but for the
manganites the on-site Coulomb interaction is strong and thus of
particular importance.  Before describing the many-electron states of
manganese ions in detail, let us review some of the basic facts for
the Coulomb interaction of $d$ electrons. Expanding the Coulomb
interaction between two charges in terms of spherical harmonics,
\begin{equation}
   \fl\quad V = \frac{e^2}{|\vec{r}_1 - \vec{r}_2|} 
    = \frac{e^2}{r_>}\sum_{k=0}^{\infty}\left(\frac{r_<}{r_>}\right)^k
    \frac{4\pi}{2k+1} \sum_{p=-k}^{k} Y_{k p}(\theta_1,\phi_1) Y_{k
      p}^{*}(\theta_2,\phi_2)
\end{equation}
with $r_> = \max(r_1,r_2)$ and $r_< = \min(r_1,r_2)$, we find that the
matrix element between different single electron functions
$\psi_\alpha = R_{n_\alpha l_\alpha} Y_{l_\alpha
  m_\alpha}s_{\sigma_\alpha}$ is given by
\begin{eqnarray}
  \fl \langle\psi_\alpha\psi_\beta|V|\psi_\gamma\psi_\delta\rangle & = &
  \sum_{k=0}^{\infty} F^k_{\alpha\beta\gamma\delta}\,
  c^k_{l_\alpha m_\alpha;l_\gamma m_\gamma}\,
  c^k_{l_\beta m_\beta;l_\delta m_\delta}\,
  \delta_{m_\alpha+m_\beta,m_\gamma+m_\delta}\,
  \delta_{\sigma_\alpha,\sigma_\gamma}\delta_{\sigma_\beta,\sigma_\delta}\,,
  \nonumber{}\\
  \fl F^k_{\alpha\beta\gamma\delta} & = &
   \langle R_{n_\alpha l_\alpha}R_{n_\beta l_\beta}|
  \frac{e^2 r_<^k}{r_>^{k+1}} |R_{n_\gamma l_\gamma}R_{n_\delta
    l_\delta}\rangle\,,\\
  \fl c^k_{lm;l'm'} & = & \sqrt{\frac{4\pi}{2k+1}}\ 
  \langle Y_{l m}|Y_{k m-m'} Y_{l' m'}\rangle\,.\nonumber{}
\end{eqnarray}
Here we already made use of the fact that the Coulomb interaction
preserves the $z$ component of the angular momentum. However, the
special structure of the matrix elements $c^k_{lm;l'm'}$ imposes
further restrictions on the expansion index $k$ and suggests a close
relation to Clebsch Gordan coefficients. Indeed, long ago
Racah~\cite{Ra42b} derived
\begin{eqnarray}
  \fl c^k_{lm;l'm'} & = & (-1)^{g-l} \sqrt{\frac{1}{2}(2l'+1)C_{lkl'}}\
  \clebsch{l'}{m'}{k}{m-m'}{l}{m}\,,\\
  \fl C_{lkl'} & = & \cases{0 & for $l+l'+k$ odd\,,\\
    \frac{2(l+l'-k)!(l+k-l')!(l'+k-l)!g!^2
    }{(l+l'+k+1)!(g-l)!^2(g-l')!^2(g-k)!^2} & for $2g := l+l'+k$ even\,,}
  \nonumber{}
\end{eqnarray}
which leads to the conditions of even $l+l'+k$ and $|l-l'| \le k \le
l+l'$. For the present case of $3d$ electrons the values of $k$ are
thus restricted to $k=0,2,4$, and the Coulomb matrix elements reduce
to a sum of only {\em three} terms,
\begin{equation}
  \fl \langle\psi_\alpha^{3d}\psi_\beta^{3d}|V
  |\psi_\gamma^{3d}\psi_\delta^{3d}\rangle = 
  \sum_{\kappa=0}^{2}
  F^{2\kappa}\,
  c^{2\kappa}_{2,m_\alpha;2,m_\gamma}\,
  c^{2\kappa}_{2,m_\delta;2,m_\beta}\,
  \delta_{m_\alpha+m_\beta,m_\gamma+m_\delta}\,
  \delta_{\sigma_\alpha,\sigma_\gamma}\delta_{\sigma_\beta,\sigma_\delta}\,.
\end{equation}
In realistic situations the integrals over the radial parts of the 
wave functions, 
\begin{equation}
  F^k = \langle R_{32}(r_1) R_{32}(r_2)|
  \frac{e^2r_<^k}{r_>^{k+1}}|R_{32}(r_1) R_{32}(r_2)\rangle\,,
\end{equation}
are usually hard to calculate exactly and are therefore taken as free
parameters which depend on details of the considered ions and
compounds. Instead of the above $F^k$ the so-called Racah parameters,
\begin{equation}
  A = F^0 - F^4/9,\quad B = \frac{9F^2-5F^4}{441},\quad C = \frac{5F^4}{63}\,,
\end{equation}
are more common, because they simplify the notation of the Coulomb
matrix elements. Since the $F^k$ are positive and monotonously
decreasing in $k$~\cite{CS35,Gr71}, the Racah parameters are positive
as well. In addition, for many transition metals the ratio $C/B$ is
almost constant and of the order of $4$ to $5$. Later we make use of
this by expressing the Coulomb interaction in terms of just two
parameters $U$ and $J_h$. In \Tref{tabracah} we show estimates of $A$,
$B$ and $C$ for manganese obtained from spectroscopic data.

\begin{table} 
  \caption{\label{tabracah}Different estimates for the Racah parameters 
    of manganese ions (in eV) and corresponding references.}
  \begin{indented} 
  \item[]\begin{tabular}{ccccccc} 
      \br
      ion & $A$ & $B$ & $C$ & reference\\
      \mr 
      Mn$^{2+}$ & 6.05 & 0.107 & 0.477 & \cite{BMSNF92,TS54b}\\
      & 5.43 & 0.1190 & 0.4122 & \cite{ZS90}\\
      Mn$^{3+}$ & 6.40 & 0.120 & 0.552 & \cite{BMSNF92,TS54b}\\
      & 5.26 & 0.120 & 0.552 & \cite{MF95,TS54b}\\
      Mn$^{4+}$ & 6.58 & 0.132 & 0.610 & \cite{BMSNF92,TS54b}\\
      \br
    \end{tabular}
  \end{indented} 
\end{table}

Based on the above introduction we are now in the position to
construct many-electron eigenstates of the Coulomb interaction for a
single ion in a cubic crystal field, which define the starting point
of our modelling. Neglecting spin-orbit coupling, which is usually
very small for compounds of the first transition metal series, the
Coulomb matrix can be decomposed into blocks of given spin and cubic
symmetry. Especially for the low-energy states it turns out that this
symmetrisation diagonalises the Coulomb matrix, i.e. we immediately
find the ground-states of the $n$ electron systems. Taking into
account the crystal field splitting, for the Mn$^{4+}$ ion ($d^3$) the
ground state is obtained by triply occupying the $t_{2g}$ levels and
forming a state of maximum spin (Hunds rule). This leads to the spin
quartet and orbital singlet $\spterm{4}{\!A}{2}$, which in terms of
the fermionic creation operators $c^{\dagger}_{\nu\sigma}$ reads
\begin{equation}\label{base_A2}
  |a_2,\frac{3}{2},m\rangle =
  \sqrt{\frac{(\frac{3}{2}-m)!}{3!(\frac{3}{2}+m)!}}\,
  (S^{+})^{(\frac{3}{2}+m)}\,
  c^{\dagger}_{\xi\downarrow} c^{\dagger}_{\eta\downarrow}
  c^{\dagger}_{\zeta\downarrow}|0\rangle\,,
\end{equation}
and its Coulomb energy is $\epsilon(\spterm{4}{\!A}{2}) = 3A-15B$. For
Mn$^{3+}$ ($d^4$) the ground state is a spin quintet and orbital
doublet $\spterm{5}{E}{}$ with components
\begin{eqnarray}\label{base_E}
  |\theta,2,m\rangle & = &
  + \,\sqrt{\frac{(2-m)!}{4!(2+m)!}}\,(S^{+})^{(2+m)}\,
  c^{\dagger}_{\varepsilon\downarrow} c^{\dagger}_{\xi\downarrow}
  c^{\dagger}_{\eta\downarrow} c^{\dagger}_{\zeta\downarrow}|0\rangle\,,\\
  |\varepsilon,2,m\rangle & = &
  - \,\sqrt{\frac{(2-m)!}{4!(2+m)!}}\,(S^{+})^{(2+m)}\,
  c^{\dagger}_{\theta\downarrow} c^{\dagger}_{\xi\downarrow}
  c^{\dagger}_{\eta\downarrow} c^{\dagger}_{\zeta\downarrow}|0\rangle\,,
\end{eqnarray}
which reflects the freedom of choosing one of the $e_{g}$ levels when
adding another electron to the $d^3$ system and requiring strong Hunds
rule coupling, i.e. maximal spin. The Coulomb energy of the
$\spterm{5}{E}{}$ state is $\epsilon(\spterm{5}{E}{}) = 6A-21B$. Note
also that in the $\theta$ component of the many-electron state the
$\varepsilon$ single-electron level is occupied, and vice versa, which
is caused by the $d^3$ state belonging to $A_2$.

\section{Perturbation theory in the electronic hopping}
\begin{figure}
  \begin{center}
    \includegraphics[width = 0.5\linewidth]{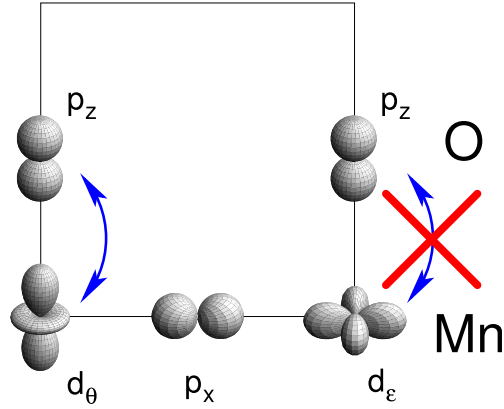}
  \end{center}
  \caption{Illustration of the orbital anisotropy of the electronic 
    hopping via bridging oxygen sites.}
  \label{fighoppz}
\end{figure}

The energy difference between the above ionic ground states and the
corresponding lowest excitations is given by the crystal field
splitting $\Delta_{\rm cf}$, or by a Coulomb energy of at least
$8B+3C$.
Both energy scales are large compared to the hopping matrix element of
$e_g$ electrons between two neighbouring manganese sites, which is of
the order of $t\approx0.3$ to $0.4$~eV~\cite{Quea98,Saea96,FO99}. The
hopping of $t_{2g}$ electrons, $t_{\pi}$, is even smaller. The
derivation of an effective low-energy Hamiltonian can thus be based on
a perturbative expansion in terms of $t_{(\pi)}$. Since the transfer
of electrons between different manganese sites always proceeds via the
completely filled $2p$ shell of bridging oxygen sites, the hopping
matrix elements acquire a particular orbital dependence, which leads
to effective spin-orbital interactions and finally causes all the
interesting patterns of orbital ordering or disorder observed in the
manganites.  As \Fref{fighoppz} illustrates for bonds in
$z$-direction, the $\varepsilon$ orbital by symmetry has no overlap
between any of the three oxygen $p$ orbitals, whereas for the $\theta$
orbital only the overlap to $p_z$ is finite. Similarly we find that
the $t_{2g}$ orbitals have finite overlap at most to one of $p_x$ or
$p_y$, and we can thus summarise all possible hopping processes in
$z$-direction with the Hamiltonian
\begin{equation}\label{hamhoppz}
\fl \quad H_{\rm hopp}^z = -\sum_{i,\sigma} \left[
    t\,c^{\dagger}_{i,\theta\sigma} c^{}_{i+z,\theta\sigma}
    +t_{\pi} (c^{\dagger}_{i,\xi\sigma} c^{}_{i+z,\xi\sigma} +
    c^{\dagger}_{i,\eta\sigma} c^{}_{i+z,\eta\sigma})
  + {\rm H.c.}\right]
\end{equation}
At first glance the situation for $x$ and $y$-bonds appears to be more
involved. However, using the cubic point symmetry operations, e.g. the
diagonal rotation $C_3^d$, we can easily derive the corresponding
hopping matrix elements by simply inserting the rotated orbitals into
\Eref{hamhoppz}. More formally we introduce the operators
\begin{eqnarray}\label{defRdelta}
    R_x & = (C_3^d)^{1}\,,\ R_y = (C_3^d)^{2} = (C_3^d)^{-1}\,,\
    R_z = (C_3^d)^{3} = 1\,,\\
    C_3^d & :\ c^{}_{\theta/\varepsilon}\rightarrow 
    -(c^{}_{\theta/\varepsilon} \mp \sqrt{3}c^{}_{\varepsilon/\theta})/2\,;\ 
    c^{}_{\xi/\eta/\zeta} \rightarrow c^{}_{\eta/\zeta/\xi}\,,\nonumber{}
\end{eqnarray}
and obtain for the complete hopping Hamiltonian
\begin{equation}\label{hamhopp}
\fl \quad H_{\rm hopp} =
  -\sum_{i,\delta,\sigma} R_\delta\left[t\,c^{\dagger}_{i,\theta\sigma}
    c^{}_{i+\delta,\theta\sigma} +
    t_{\pi} (c^{\dagger}_{i,\xi\sigma} c^{}_{i+\delta,\xi\sigma} +
    c^{\dagger}_{i,\eta\sigma} c^{}_{i+\delta,\eta\sigma})
  + {\rm H.c.}\right]
\end{equation}

In general $H_{\rm hopp}$ connects the ionic ground state at one site
to a large number of Coulomb excited states at the other site. A
complete description of the electronic subsystem would thus require a
huge local Hilbert space, which includes basically all ionic
excitations. However, these high-energy excitations are irrelevant for
most of the features we want to model, and we thus restrict the local
Hilbert space to the ionic ground states derived in \Sref{sec_ion} and
account for the excitations in the framework of standard degenerate
perturbation theory.
Then the first order of the effective Hamiltonian is given by the
matrix elements of $H_{\rm hopp}$ between basis states of the form
(\ref{base_A2}) and (\ref{base_E}) on neighbouring sites. These basis
states differ in the total on-site spin $|\vec{S}_i|$. To achieve an
uniform description it is thus convenient to rewrite the spin degrees
of freedom in terms of Schwinger bosons~\cite{MaI88}, which provide a
simple means for changing $|\vec{S}_i|$,
\begin{eqnarray}
  S_i^+ = a_{i,\uparrow}^{\dagger} a_{i,\downarrow}^{}\,, \qquad & 
  S_i^z = (a_{i,\uparrow}^{\dagger} a_{i,\uparrow}^{} -
  a_{i,\downarrow}^{\dagger} a_{i,\downarrow}^{})/2\,, \nonumber{}\\
  S_i^-  = a_{i,\downarrow}^{\dagger} a_{i,\uparrow}^{}\,, \qquad & 
  |\vec{S}_i| = (a_{i,\uparrow}^{\dagger} a_{i,\uparrow}^{} 
  + a_{i,\downarrow}^{\dagger}a_{i,\downarrow}^{})/2\,.
\end{eqnarray}
In addition, we can also reduce the number of fermionic operators and
introduce hole operators $d_{i,\theta}^{(\dagger)}$ and
$d_{i,\varepsilon}^{(\dagger)}$ instead. The orbital part of our
basis states and corresponding projection operators then read
\begin{eqnarray}
  |\theta\rangle
  = d^{\dagger}_{\theta}|0\rangle\,, \qquad  &
  |\varepsilon\rangle
  = d^{\dagger}_{\varepsilon} |0\rangle\,, \qquad &
  |a_2\rangle 
  = d^{\dagger}_{\theta} d^{\dagger}_{\varepsilon}
  |0\rangle\,, \nonumber{}\\
  P^{\theta}_{}  = n^{}_{\theta}(1-n^{}_{\varepsilon})\,, \qquad &
  P^{\varepsilon}_{}  = n^{}_{\varepsilon}(1-n^{}_{\theta})\,, \qquad  &
  P^{a_2}_{}  = n^{}_{\varepsilon} n^{}_{\theta}\,,
\end{eqnarray}
where the vacuum $|0\rangle$ is invariant under cubic symmetry
operations, i.e. belongs to $A_1$. Calculating the effective hopping
matrix element for a $z$-bond we are mainly concerned with the spin
part, whereas the orbital part is rather trivial, since only $\theta$
electrons are allowed to tunnel. Given the spin-$2$ state
$|\varepsilon,2,m\rangle$ on site $i$ and the spin-$3/2$ state
$|a_2,3/2,m'\rangle$ on site $j$ the hopping of a $\theta$ electron
will transfer a spin-$1/2$ from $i$ to $j$. Projecting the new states
onto our basis will lead to $|a_2\rangle$ at $i$ and
$|\varepsilon\rangle$ at $j$. In addition, the hopping conserves the
total bond spin $\vec{S}_i + \vec{S}_j$. Combining all three
properties of the process, i.e. transfer of a spin-$1/2$, increase of
$|\vec{S}_j|$ and lowering of $|\vec{S}_i|$, as well as conservation
of $\vec{S}_i + \vec{S}_j$, in terms of Schwinger bosons we can almost
guess the effective matrix element to be proportional to
$a_{j,\uparrow}^{\dagger} a_{i,\uparrow}^{} +
a_{j,\downarrow}^{\dagger} a_{i,\downarrow}^{}$. Indeed, a more
careful calculation~\cite{WLF01a} of the
double-exchange~\cite{Ze51b,AH55} matrix element shows that
corresponding prefactor is $-t/(2S+1)$, where $S$ is the amplitude of
the shorter of the two on-site spins. For a bond in $z$-direction the
first order of the effective Hamiltonian is thus given by
\begin{equation}\label{hamel1z}
  H^{1,z}_{i,j}  =
  -\frac{t}{4}\left(a^{}_{i,\uparrow} a^{\dagger}_{j,\uparrow}
    +a^{}_{i,\downarrow}a^{\dagger}_{j,\downarrow}\right)\
  d^{\dagger}_{i,\theta} n^{}_{i,\varepsilon}
  d^{}_{j,\theta} n^{}_{j,\varepsilon} + {\rm H.c.}
\end{equation}
Note that the amplitude of the local spin is restricted by $2
|\vec{S}_i| = a^{\dagger}_{i,\uparrow} a^{}_{i,\uparrow} +
a^{\dagger}_{i,\uparrow} a^{}_{i,\uparrow} = 5 -
(n^{}_{i,\varepsilon}+n^{}_{i,\theta})$.

\begin{table} 
  \caption{\label{tabexcit}Excitations of a single manganese ion that can 
    be reached by adding or subtracting an $e_g$ or $t_{2g}$ electron to 
    the basis states $\spterm{5}{E}{}$ and $\spterm{4}{\!A}{2}$. Starred 
    Coulomb energies in the last column are approximate (see text and 
    Ref.~\cite{Gr71}).}
  \begin{indented} 
  \item[]\renewcommand{\arraystretch}{1.4}\begin{tabular}{ccccc} 
      \br 
      initial state & \multicolumn{2}{c}{operation} & final state & energy\\
      \mr
      $\spterm{5}{E}{}[t_2^3(\spterm{4}{\!A}{2})e]$&$d^4\nearrow d^5$
      &$c_{e\downarrow}^{\dagger}$&
      $\spterm{6}{\!A}{1}[t_2^3(\spterm{4}{\!A}{2})e^2(\spterm{3}{\!A}{2})]$&
      $10A-35B$\\
      & &$c_{e\uparrow}^{\dagger}$&
      $\spterm{4}{\!A}{1}[t_2^3(\spterm{4}{\!A}{2})e^2(\spterm{3}{\!A}{2})]$&
      $10A-25B+5C$\\
      & & &
      $\spterm{4}{\!A}{2}[t_2^3(\spterm{4}{\!A}{2})e^2(\spterm{1}{\!A}{1})]$&
      $10A-13B+7C$\\
      & & &
      $\spterm{4}{E}{}[t_2^3(\spterm{4}{\!A}{2})e^2(\spterm{1}{E}{})]$&
      $10A-21B+5C^{*}$\\
      & &$c_{t_2\uparrow}^{\dagger}$&
      $\spterm{4}{T}{1}[t_2^4(\spterm{3}{T}{1})e]$&
      $10A-25B+6C^{*}$\\
      & & &
      $\spterm{4}{T}{2}[t_2^4(\spterm{3}{T}{1})e]$&
      $10A-17B+6C^{*}$\\
      \cline{2-5}
      &$d^4\searrow d^3$&$c_{e\downarrow}^{}$&
      $\spterm{4}{\!A}{2}[t_2^3]$&
      $3A-15B$\\
      & &$c_{t_2\downarrow}^{}$&
      $\spterm{4}{T}{1}[t_2^2(\spterm{3}{T}{1})e]$&
      $3A-3B^{*}$\\
      & & &
      $\spterm{4}{T}{2}[t_2^2(\spterm{3}{T}{1})e]$&
      $3A-15B$\\
      \mr
      $\spterm{4}{\!A}{2}[t_2^3]$&$d^3\nearrow d^4$
      &$c_{e\downarrow}^{\dagger}$&
      $\spterm{5}{E}{}[t_2^3(\spterm{4}{\!A}{2})e]$&
      $6A-21B$\\
      & &$c_{e\uparrow}^{\dagger}$&
      $\spterm{3}{E}{}[t_2^3(\spterm{4}{\!A}{2})e]$&
      $6A-13B+4C^{*}$\\
      & &$c_{t_2\uparrow}^{\dagger}$&
      $\spterm{3}{T}{1}[t_2^4]$&
      $6A-15B+5C^{*}$\\
      \cline{2-5}
      &$d^3\searrow d^2$&$c_{t_2\downarrow}^{}$&
      $\spterm{3}{T}{1}[t_2^2]$&
      $A-5B^{*}$\\
      \br
    \end{tabular}\renewcommand{\arraystretch}{1.0}
  \end{indented} 
\end{table}
\begin{table} 
  \caption{\label{tabmatelem}The matrix elements 
    $\langle d^n d^m | H_{\rm hopp}^z | d^{n-1} d^{m+1}\rangle = 
    O_{\alpha\beta}\, S(x)\, t_{(\pi)}$ together with the corresponding 
    excitation energies $\Delta\epsilon$. For brevity we denote vanishing 
    matrix elements with dots, write $x=\vec{S}_i \vec{S}_j$ and give 
    the squared $S(x)$. }
  \begin{indented} 
  \item[]\renewcommand{\arraystretch}{1.2}\begin{tabular}{ccccccc} 
      \br 
      \multicolumn{4}{c}{$O_{\alpha\beta}$}&$S(x)^2$&$t_{(\pi)}$&$\Delta\epsilon$\\
      $\theta\theta$&$\theta\varepsilon$&$\varepsilon\theta$&$\varepsilon\varepsilon$&&&\\
      \mr
      $\cdot$&$\cdot$&$-1$&$\cdot$&$\frac{3}{5}\frac{4-x}{16}$&$t$&$A+2B+5C$\\
      $\cdot$&$\cdot$&$\cdot$&$-1$&$\frac{4-x}{16}$&$t$&$A+14B+7C$\\
      $\cdot$&$\cdot$&$1$&$\cdot$&$\frac{4-x}{16}$&$t$&$A+6B+5C$\\
      $\cdot$&$\cdot$&$\cdot$&$-1$&$\frac{4-x}{16}$&$t$&$A+6B+5C$\\
      $\cdot$&$\cdot$&$-1$&$\cdot$&$\frac{6+x}{10}$&$t$&$A-8B$\\
      \mr
      $-\frac{3}{4}$&$\pm\frac{\sqrt{3}}{4}$&$\pm\frac{\sqrt{3}}{4}$&$-\frac{1}{4}$&$\frac{4-x}{8}$&$t_{\pi}$&$A+14B+6C$\\
      $\frac{1}{4}$&$\pm\frac{\sqrt{3}}{4}$&$\pm\frac{\sqrt{3}}{4}$&$\frac{3}{4}$&$\frac{4-x}{8}$&$t_{\pi}$&$A+10B+6C$\\
      $\mp\frac{\sqrt{3}}{4}$&$\frac{3}{4}$&$-\frac{1}{4}$&$\pm\frac{\sqrt{3}}{4}$&$\frac{4-x}{8}$&$t_{\pi}$&$A+22B+6C$\\
      $\pm\frac{\sqrt{3}}{4}$&$\frac{1}{4}$&$-\frac{3}{4}$&$\mp\frac{\sqrt{3}}{4}$&$\frac{4-x}{8}$&$t_{\pi}$&$A+2B+6C$\\
      \br 
      $\theta a_2$&$\varepsilon a_2$&$a_2\theta$&$a_2\varepsilon$&&&\\
      \mr
      $\cdot$&$1$&$\cdot$&$\cdot$&$\frac{3-x}{8}$&$t$&$8B+4C$\\
      \mr
      $\pm\frac{\sqrt{3}}{2}$&$-\frac{1}{2}$&$\cdot$&$\cdot$&$\frac{3-x}{6}$&$t_{\pi}$&$18B+5C$\\
      $\frac{1}{2}$&$\pm\frac{\sqrt{3}}{2}$&$\cdot$&$\cdot$&$\frac{3-x}{6}$&$t_{\pi}$&$6B+5C$\\
      \mr
      $\cdot$&$\cdot$&$\pm\frac{\sqrt{3}}{2}$&$-\frac{1}{2}$&$\frac{3-x}{6}$&$t_{\pi}$&$2A+6B+6C$\\
      $\cdot$&$\cdot$&$-\frac{1}{2}$&$\pm\frac{\sqrt{3}}{2}$&$\frac{3-x}{6}$&$t_{\pi}$&$2A+14B+6C$\\
      \br 
      &&&$a_2 a_2$&&&\\
      \mr
      &&&$1$&$\frac{9-4x}{9}$&$t_{\pi}$&$A+10B+5C$\\
      \br
    \end{tabular}\renewcommand{\arraystretch}{1.0}
  \end{indented} 
\end{table}

The second order of perturbation theory in $t_{(\pi)}$ is a bit more
involved, since now all virtual excitations of the basis states have
to be taken into account. In \Tref{tabexcit} we list all eleven
many-particle states that can be constructed by adding or subtracting
one $e_g$ or $t_{2g}$ electron to the basis states from
Equations~(\ref{base_A2}) and~(\ref{base_E}), thereby respecting cubic
and spin symmetry. Since usually there are other configurations 
belonging to the same representations of orbital and spin symmetry,
these states are not necessarily eigenstates of the on-site Coulomb
interaction. However, since these other states have no overlap to our
basis, taking into account the exact Coulomb eigenstates will only
slightly modify the energy denominator of the second order terms, but
not the general structure of the matrix elements. It is thus
sufficient to consider the expectation value of the Coulomb energy in
the above configurations, and the corresponding approximate energies
are marked with a star in \Tref{tabexcit}. Proceeding further, we can
now place all pairs of the three basis states on a single Mn-Mn bond
and form states of given total bond spin $\vec{S}_b$. Then we need to
calculate the overlap to all pairs of excited states that are
connected to our bond ground-state by a single electron transfer and
share the same bond spin $\vec{S}_b$. Clearly, there is a reasonable
number of different combinations, and the use of some computer algebra
system is recommended. Each of the matrix elements
\begin{equation}
  \langle d^nd^m|H_{\rm hopp}|d^{n-1}d^{m+1}\rangle
  = O_{\alpha\beta}\, S(\vec{S}_i\vec{S}_j)\, t_{(\pi)}\,,
  \quad\alpha,\beta\in\{\theta,\varepsilon,a_2\}
\end{equation}
can be decomposed into an orbital part $O_{\alpha\beta}$, a spin part
$S(\vec{S}_i\vec{S}_j)$ and the corresponding transfer amplitude
$t_{(\pi)}$.  In \Tref{tabmatelem} we list these components for all
$d^4d^4\rightleftharpoons d^3d^5$, $d^4d^3 \rightleftharpoons d^3d^4$,
$d^3d^4 \rightleftharpoons d^2d^5$ and $d^3d^3 \rightleftharpoons
d^2d^4$ excitations for a bond in $z$-direction, i.e. $j=i+z$.
Comparing the orbital parts $O_{\alpha\beta}$ with the rotation
properties of the $\theta$ and $\varepsilon$ functions given in
\Tref{tabirrep}, we observe that the $O_{\alpha\beta}$ can be
simplified, if we change the orbital quantisation axis from $z$ to $x$
or $y$. For instance, the four matrix elements $-\{3, \sqrt{3},
\sqrt{3}, 1\}/4$ in line 6 of \Tref{tabmatelem}, which connect the
functions $\{\theta\theta, \theta\varepsilon, \varepsilon\theta,
\varepsilon\varepsilon\}$ to the appropriate excitation, reduce to a
single matrix element $-1$ originating from the bond state
$\varepsilon_x\varepsilon_x$ with $\varepsilon_x = -(\sqrt{3}\,\theta +
\varepsilon)/2$.  Hereafter we use these properties to express some of
the interactions with the help of the rotation operators $R_\delta$,
$\delta\in\{x,y,z\}$.

Putting together all the above pieces we are now in the position to
formulate the second order contributions to our effective electronic
Hamiltonian.  For simplicity, we neglect all terms which involve
hopping processes between three sites, and restrict ourselves to those
terms, where an electron hops back and forth on a single Mn-Mn bond,
\begin{equation}
  H = -\sum_{{\langle ij\rangle}\atop{
      \alpha_i\alpha_j\beta_i\beta_j,\Psi}}
  \frac{|\alpha_i\alpha_j\rangle\langle\alpha_i\alpha_j|
    \,H_{\rm hopp}\,|\Psi\rangle\langle\Psi|\,H_{\rm hopp}\,
    |\beta_i\beta_j\rangle\langle \beta_i\beta_j|}{\Delta\epsilon(\Psi)}\,.
\end{equation}
In addition, we follow Ref.~\cite{FO99} by assuming $C\approx 4B$ and
introducing new Coulomb parameters $U$ and $J_h$, which refer to a
Hubbard like on-site repulsion and a Hunds rule coupling,
respectively. The actual choice of the relation between $U$, $J_h$ and
$A$, $B$, $C = 4B$ is a matter of convention. Having in mind the
undoped compounds and a bond with $d^4(\spterm{5}{E}{})$ basis state
on each site, we may ascribe $U$ to the minimal energy required for
transferring one of the $e_g$ electrons and forming a low spin $d^5$
state on one of the sites. On the other hand, the excitation with a
high-spin $d^5$ state should have energy $U - 5J_h$, where the factor
$5$ reminds of the energy difference originating from an exchange term
$J_h\vec{S}\vec{s}$ with $|\vec{S}|=2$ and $|\vec{s}|=1/2$. This
choice leads to
\begin{eqnarray}
  U & = \epsilon[d^5(\spterm{4}{\!A}{1}) d^3(\spterm{4}{\!A}{2})]
  - \epsilon[d^4(\spterm{5}{E}{})d^4(\spterm{5}{E}{})] = A+2B+5C\,,
  \nonumber{}\\
  5 J_h & = \epsilon[d^5(\spterm{4}{\!A}{1}) d^3(\spterm{4}{\!A}{2})] -
  \epsilon[d^5(\spterm{6}{\!A}{1})d^3(\spterm{4}{\!A}{2})] = 10B+5C\,,
\end{eqnarray}
and after some algebra for a $z$-bond we obtain the following
contribution of second order in $t_{(\pi)}$
\begin{eqnarray}\label{hamelz}
\fl  H^{2,z}_{i,j} & =  t^2\,\frac{\vec{S}_{i}\vec{S}_{j}-4}{8}
  \left[\frac{(4U+J_h)\,P^{\varepsilon}_{i}P^{\theta}_{j}} {5 U
      (U+\frac{2}{3} J_h)} +
    \frac{(U+2J_h)\,P^{\varepsilon}_{i}P^{\varepsilon}_{j}}
    {(U+\frac{10}{3} J_h)(U+\frac{2}{3} J_h)}\right]\nonumber{}\\
\fl  & - t^2\,\frac{\vec{S}_{i} \vec{S}_{j}+6}{10(U-5J_h)}
  \,P^{\varepsilon}_{i}P^{\theta}_{j} + t_{\pi}^2\,\frac{\vec{S}_{i}
    \vec{S}_{j}-4}{8}
  \left[\frac{R_{x}(P^{\varepsilon}_{i}P^{\varepsilon}_{j})+
      R_{y}(P^{\varepsilon}_{i}P^{\varepsilon}_{j})}{U+8J_h/3}\right.\nonumber{}\\
\fl  &\ \qquad\left. + \frac{R_{x}(P^{\theta}_{i}P^{\theta}_{j})+
      R_{y}(P^{\theta}_{i}P^{\theta}_{j})}{U+2J_h} +
    \frac{(2U+\frac{14}{3}J_h)(R_{x}(P^{\varepsilon}_{i}P^{\theta}_{j})+
      R_{y}(P^{\varepsilon}_{i}P^{\theta}_{j}))}{(U+4J_h)(U+\frac{2}{3}J_h)}
  \right]\nonumber{}\\
\fl  & + t^2\,\frac{\vec{S}_{i} \vec{S}_{j}-3}{32 J_h}
  \,P^{\varepsilon}_{i}P^{a_2}_{j} + t_{\pi}^2\,
  \frac{\vec{S}_{i}\vec{S}_{j}-3}{3}
  \left[\frac{(U-2J_h)(R_{x}(P^{\varepsilon}_{i}P^{a_2}_{j})+
      R_{y}(P^{\varepsilon}_{i}P^{a_2}_{j}))}
    {\frac{19}{3}J_h (2U-\frac{7}{3}J_h)}\right.\nonumber{}\\
\fl  &\ \qquad + \left.
    \frac{(U+\frac{5}{3}J_h)(R_{x}(P^{\theta}_{i}P^{a_2}_{j})+
      R_{y}(P^{\theta}_{i}P^{a_2}_{j}))}{\frac{13}{3}J_h(2U-J_h)}
  \right]\nonumber{}\\
\fl  & + t_{\pi}^2\,\frac{\frac{4}{9}\vec{S}_{i}\vec{S}_{j}-1}{
  U+\frac{4}{3}J_h} \,P^{a_2}_{i}P^{a_2}_{j}+{\rm H.c.}
\end{eqnarray}
The first three terms reproduce the result of Ref.~\cite{FO99} for the
undoped compounds, which are characterised by a competition of one
ferromagnetic and a number of anti-ferromagnetic spin exchange terms
of order $t_{(\pi)}^2/U$.  The latter, however, are directly coupled
to pairs of orbital projectors favouring different patterns of orbital
ordering. Upon doping the ferromagnetic double-exchange interaction,
\Eref{hamel1z}, comes into play, but also anti-ferromagnetic terms of
order $t_{(\pi)}^2/J_h$ gain importance. Obviously, the complex phase
diagram and the different ordering patterns of the manganites are a
direct consequence of effects noticeable already on the level of the
microscopic Hamiltonian.

\section{Electron-phonon interaction}
\begin{figure}
  \begin{center}
    \includegraphics[width = 0.8\linewidth]{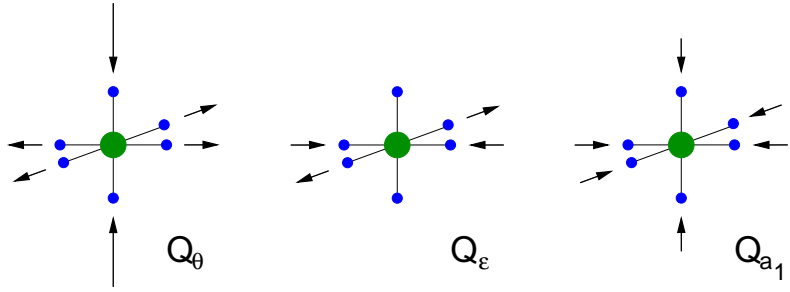}
  \end{center}
  \caption{Elongation patterns of the oxygen ions within the $E_g$ 
    and $A_{1g}$ vibration modes of a MnO$_6$ octahedron.}
  \label{figmodes}
\end{figure}
As discussed before, most undoped or weakly doped manganites show long
range (Jahn-Teller type) distortions from the ideal perovskite
structure, and changes in the magnetic or orbital ordering patterns
are usually accompanied by changes in the crystal structure. In
addition, modern experimental techniques show that besides these
cooperative static effects also dynamic and local properties of the
lattice are important for an understanding of the manganites. To name
only a few, we may think of x-ray-absorption fine-structure
data~\cite{BBKLCN98} showing a clear relationship between magnetic
order and local lattice distortion, or measurements of the atomic
pair-correlation function~\cite{LEBRB97,BPPSK00} proving the
correlation of electric resistivity and local lattice structure near
the metal-insulator transition. It is thus natural to consider
electron lattice interactions and, in particular, polaronic effects as
an important ingredient of a microscopic description of CMR
manganites~\cite{MLS95,Mi98}.

Like for the electronic, spin and orbital degrees of freedom, the
local environment of the manganese ions is the appropriate starting
point also for the modelling of the electron lattice interaction. In
\Fref{figmodes} we show the basic vibration modes of the MnO$_6$
octahedron that are even under inversion and thus susceptible for a
linear coupling to the $d$ orbitals. The modes $Q_\theta$ and
$Q_\varepsilon$, which belong to the irreducible representation $E_g$,
are responsible for the Jahn-Teller effect~\cite{JT37,Ja38}, where the
system tries to lift the degeneracy of the two $\spterm{5}{E}{}$ ionic
basis states (see \Eref{base_E}) and gain electronic energy by
lowering the point symmetry through a distortion of the lattice. As
above the electronic degrees of freedom are described by the fermionic
operators $d_{\theta / \varepsilon}^{(\dagger)}$, whereas for the
phonons we have the bosonic operators $b_{\theta /
  \varepsilon}^{(\dagger)}$. The interaction between both should be
linear in the bosons, bilinear in the fermions, and belong to the
irreducible representation $A_1$ to form a Hamiltonian that conserves
cubic symmetry. These requirements lead to one of the classical
Jahn-Teller problems, the so-called $E\otimes e$
model~\cite{LOPS58,PW84}
\begin{equation}\label{hamJT}
\fl\quad  H^{\rm JT} = g \left[
    (n^{}_{\varepsilon}-n^{}_{\theta})
    (b^{\dagger}_{\theta}+b^{}_{\theta})
  + (d^{\dagger}_{\theta} d^{}_{\varepsilon} +
    d^{\dagger}_{\varepsilon} d^{}_{\theta})
    (b^{\dagger}_{\varepsilon}+b^{}_{\varepsilon})\right]
  + \omega \left[b^{\dagger}_{\theta}b^{}_{\theta}
    + b^{\dagger}_{\varepsilon} b^{}_{\varepsilon}\right]\,,
\end{equation}
where the last term refers to the usual harmonic lattice dynamics. For
a single ion this model possesses an additional $O(2)$ symmetry, which
becomes manifest in the commutation of the Hamiltonian $H^{\rm JT}$
with the $A_2$ symmetric operator $(d^{\dagger}_{\theta}
d^{}_{\varepsilon} - d^{\dagger}_{\varepsilon} d^{}_{\theta})/2 -
(b^{\dagger}_{\theta} b^{}_{\varepsilon} - b^{\dagger}_{\varepsilon}
b^{}_{\theta})$, and leads to a two-fold degeneracy of every
eigenvalue. Whereas for a single ion this seeming contradiction to the
common notion of a Jahn-Teller system can be lifted by higher order
electron lattice interactions, for realistic compounds like the
manganites this symmetry is already broken by the orbitally
anisotropic hopping between neighbouring sites, \Eref{hamhopp}.
Unfortunately we thereby loose an advantage of this symmetry: the
simple tridiagonal structure of the $E\otimes e$ Hamiltonian
matrix~\cite{LOPS58}. Compared to a symmetric hopping, the
symmetry-breaking orbital anisotropy may also increases the
susceptibility of the system to Jahn-Teller type ordering and
polaronic effects, an issue which could be interesting for future
studies.

Coming back to the phonon modes of the MnO$_6$ octahedron, the
symmetric mode $Q_{a_1}$ should couple to a bilinear fermionic
operator belonging to the same $A_1$ representation. The most natural
operator of this type is the electronic density, and we arrive at the 
Holstein type~\cite{Ho59a,Ho59b} Hamiltonian,
\begin{equation}\label{hambreath}
  H^{\rm Hol}  = \tilde g\,
  (n^{}_{\theta} + n^{}_{\varepsilon})(b^{\dagger}_{a_1}+b^{}_{a_1})
  + \tilde\omega\ b^{\dagger}_{a_1} b^{}_{a_1}\,,
\end{equation}
which is well known from polaron physics. Since we use hole operators
instead of electrons, we prefer to adjust the expression for the local
density such that the $\spterm{5}{E}{}$ states couple to $Q_{a_1}$ but
$\spterm{4}{A}{2}$ does not, $n^{}_{i,\theta} +
n^{}_{i,\varepsilon}-2n^{}_{i,\theta}n^{}_{i,\varepsilon}$.
Generalising the above types of local electron phonon interaction to
the crystal and neglecting the weak dispersion of the optical phonons
corresponding to $Q_\theta$, $Q_\varepsilon$ and $Q_{a_1}$, we finally
arrive at the Hamiltonian
\begin{eqnarray}\label{hamep}
\fl\quad  H^{\rm EP} & = g\sum_i\left[
    (n^{}_{i,\varepsilon}-n^{}_{i,\theta})
    (b^{\dagger}_{i,\theta}+b^{}_{i,\theta}) + (d^{\dagger}_{i,\theta}
    d^{}_{i,\varepsilon} + d^{\dagger}_{i,\varepsilon}
    d^{}_{i,\theta})
    (b^{\dagger}_{i,\varepsilon}+b^{}_{i,\varepsilon})\right]\nonumber{}\\
\fl  & + \tilde g\sum_i (n^{}_{i,\theta} +
  n^{}_{i,\varepsilon}-2n^{}_{i,\theta}n^{}_{i,\varepsilon})
  (b^{\dagger}_{i,a_1}+b^{}_{i,a_1})\\
\fl  & + \omega\sum_i \left[b^{\dagger}_{i,\theta}b^{}_{i,\theta} +
    b^{\dagger}_{i,\varepsilon} b^{}_{i,\varepsilon}\right] +
  \tilde\omega\sum_i b^{\dagger}_{i,a_1}b^{}_{i,a_1}\,.\nonumber{}
\end{eqnarray}

\section{Discussion of the microscopic model and numerical results}

\begin{figure}
  \begin{center}
    \includegraphics[width=0.8\linewidth]{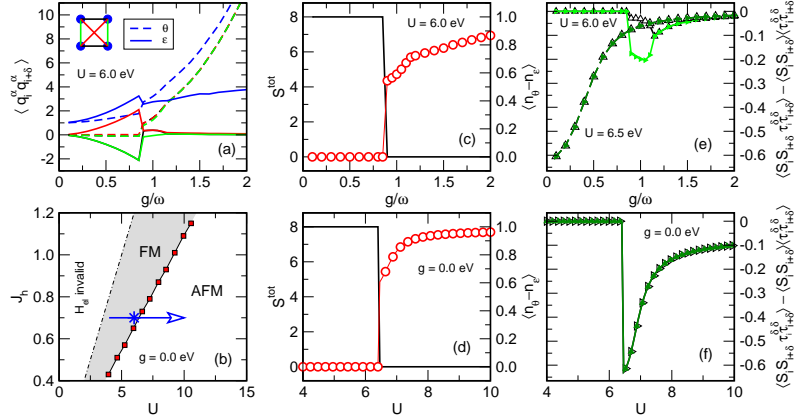}
  \end{center}
  \caption{Ground-state properties of the full model, \Eref{fullmicroham}, 
    on a four site cluster at doping $x=0$. (a) Evolution of lattice
    correlations with increasing electron phonon (EP) coupling. (b)
    Phase diagram of the model without EP coupling; the blue star and
    arrow mark parameters $U$, $J_h$ used in the other panels. (c)
    Spin and orbital ordering with increasing EP coupling $g$. (d)
    Spin and orbital ordering with increasing $U$ and $g=0$. (e)
    Decoupling of spins and orbitals with increasing EP coupling $g$.
    (f) Spin-orbital correlations at $g=0$. Throughout $J_h=0.7$~eV,
    $t=3t_\pi=0.4$~eV, $\omega = \tilde\omega=0.07$~eV, $\tilde g =
    g$; see~\cite{WF02l} for more details.}\label{figdope0}
\end{figure}

\begin{figure}
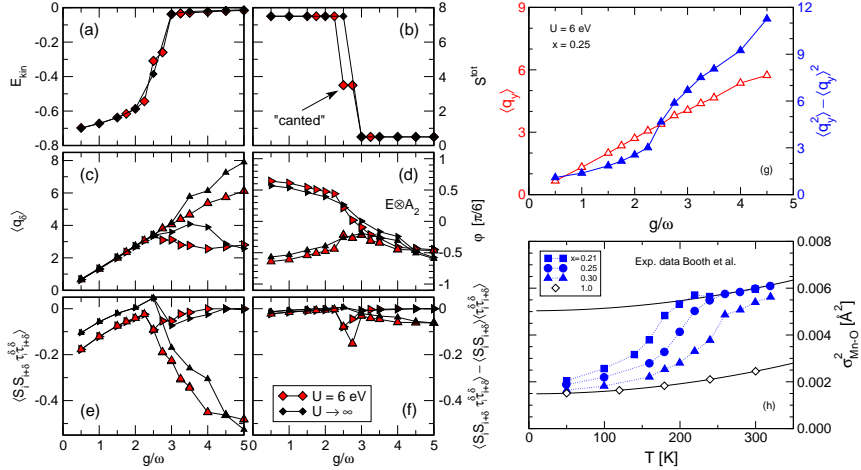

  \begin{center}
    \includegraphics[height=0.3\textheight]{figure8.eps}
    \includegraphics[height=0.3\textheight]{figure9.eps}
  \end{center}
  \caption{Ground-state properties of the full model, \Eref{fullmicroham}, 
    on a four site cluster at doping $x=0.25$. (a) Due to polaron
    formation the kinetic energy decreases with $g$. (b) At strong $g$
    double-exchange is too weak to promote ferromagnetism, i.e. the
    total spin of the clusters goes to zero. (c) Bond-lengths $\langle
    q_{x/y}\rangle$ along $x$ and $y$ direction show lattice
    distortion in the spin disordered phase. (d) The orbital
    orientation $\varphi$ in the neighbourhood of a hole is
    orbital-polaron like only for mobile carriers. (e) and (f)
    Electron-phonon interaction causes an effective decoupling of spin
    and orbital degrees of freedom. (g) and (h) For increasing $g$
    lattice fluctuations show a kink near the FM-AFM transition, quite
    similar to the experimentally observed behaviour as a function of
    temperature~\cite{BBKLCN98}. See~\cite{WF02l,WF04} for more
    details.  }\label{figdope1}
\end{figure}

\begin{figure}
  \begin{center}
    \includegraphics[width=0.8\linewidth]{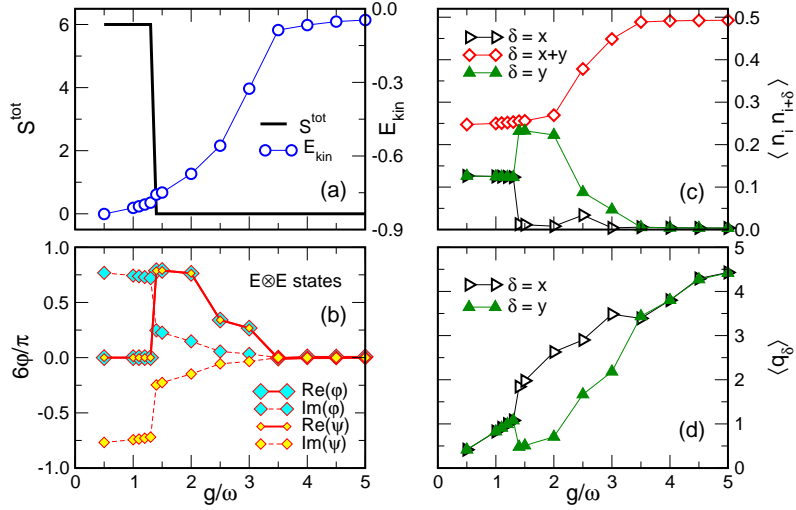}
  \end{center}
  \caption{Ground-state properties of the full model, \Eref{fullmicroham}, 
    on a four site cluster at doping $x=0.5$. (a) Similar to $x=0.25$
    increasing $g$ causes a polaronic band narrowing, but due to
    charge ordering and more important AF interactions magnetism is
    decoupled from charge dynamics. (b) For small $g$ orbital
    correlations can be characterised by complex orbital states. (c)
    Strong EP coupling leads to charge order. (d) The lattice can show
    different distortion patterns. See~\cite{WF02l} for more
    details.}\label{figdope2}
\end{figure}

\begin{figure}
  \begin{center}
    \begin{tabular}{cccc}
      & $x=0$ & $x=0.25$ & $x=0.5$\\[3mm]
      \rotatebox{90}{weak coupling} &  
      \includegraphics[width=0.2\linewidth]{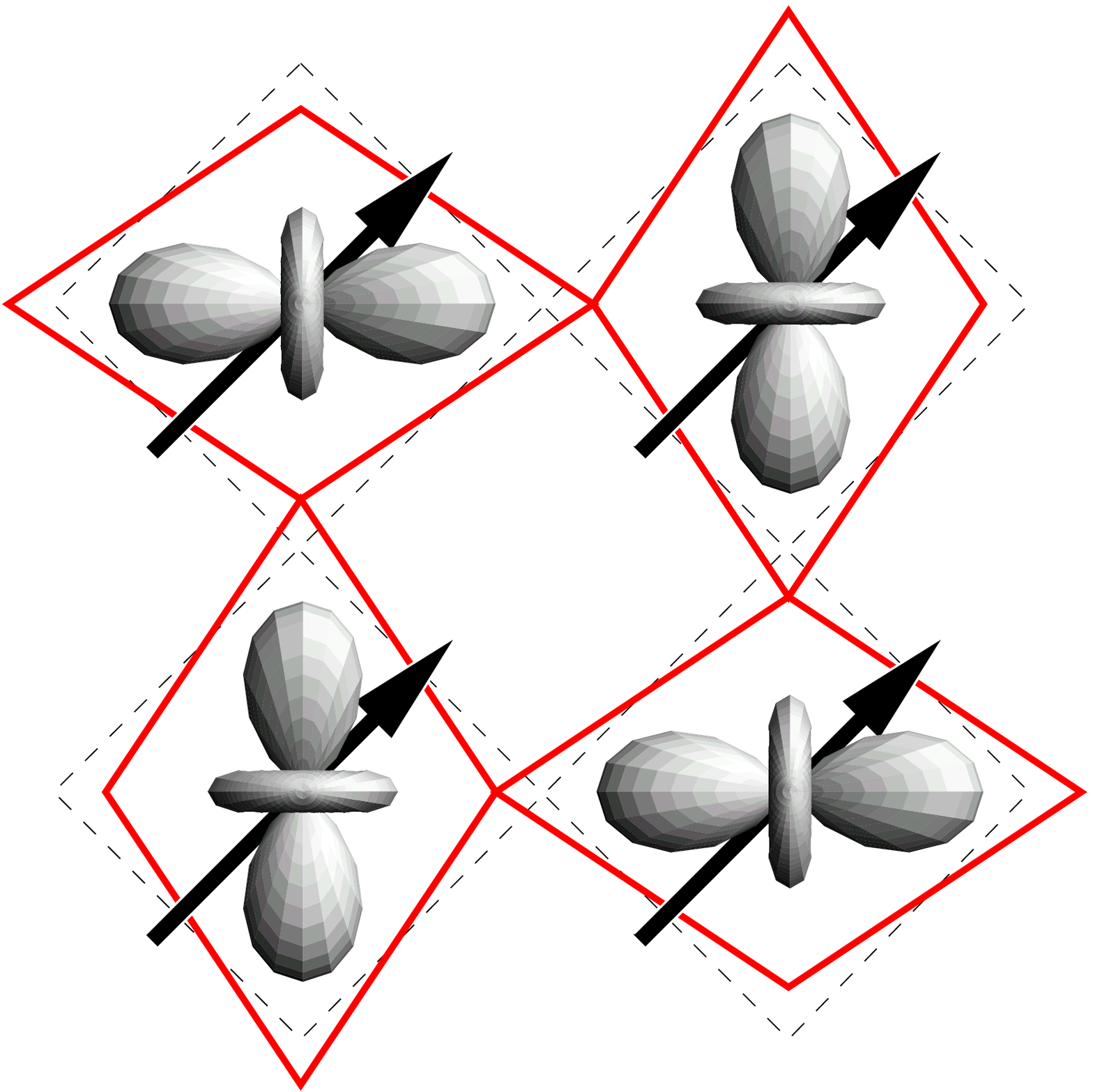} &
      \includegraphics[width=0.2\linewidth]{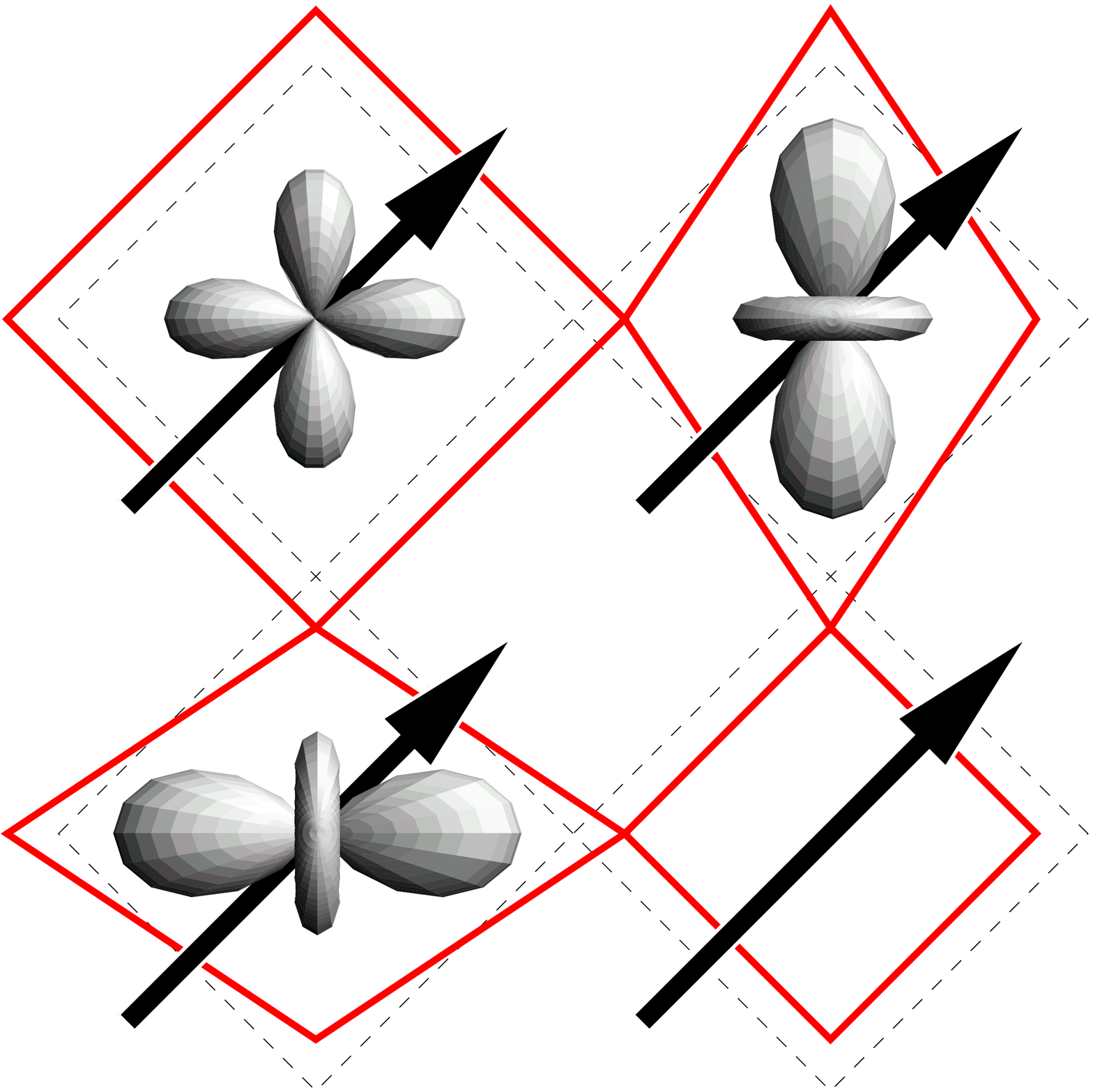} &
      \includegraphics[width=0.2\linewidth]{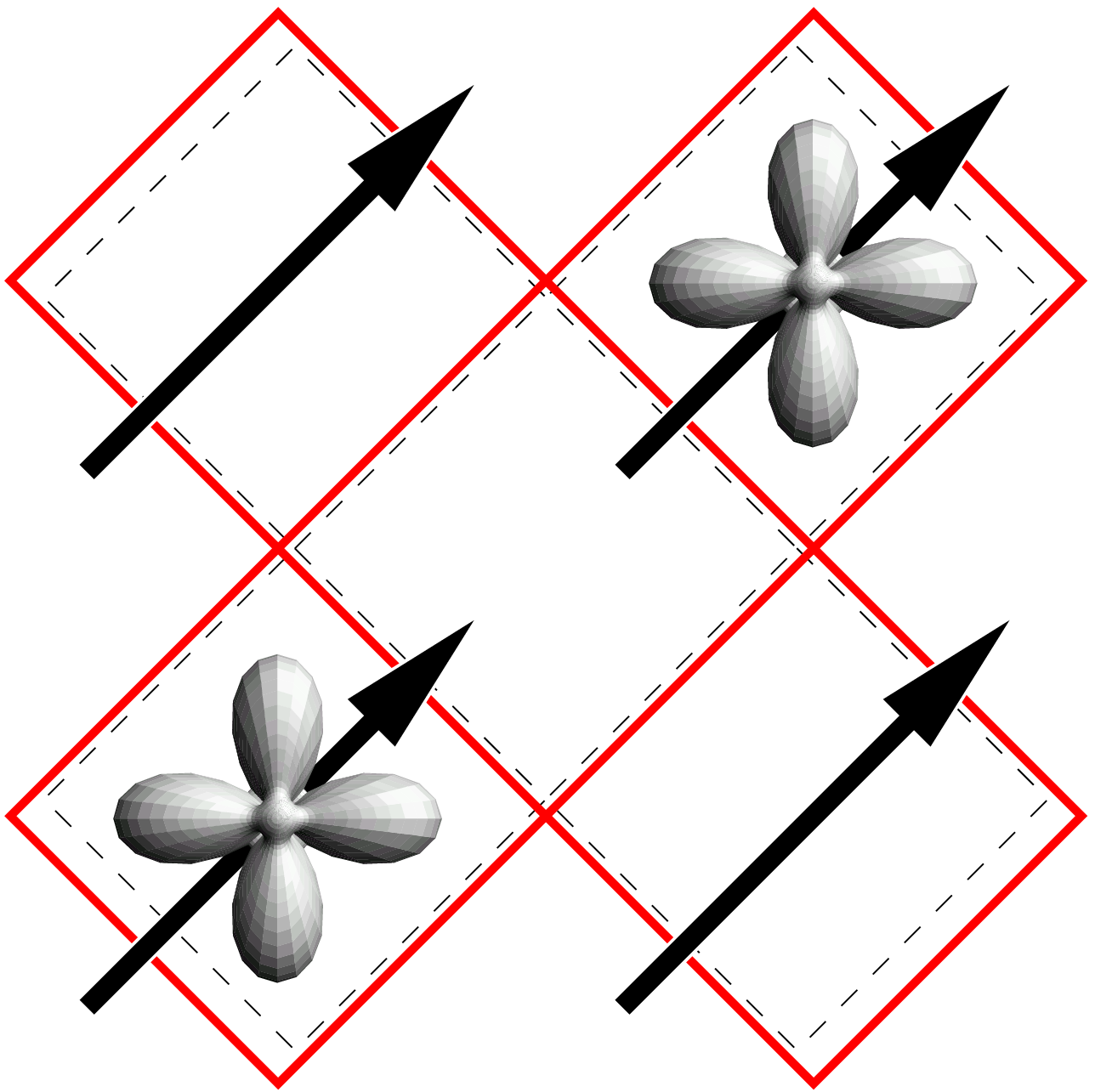} \\
      \rotatebox{90}{strong coupling} &  
      \includegraphics[width=0.2\linewidth]{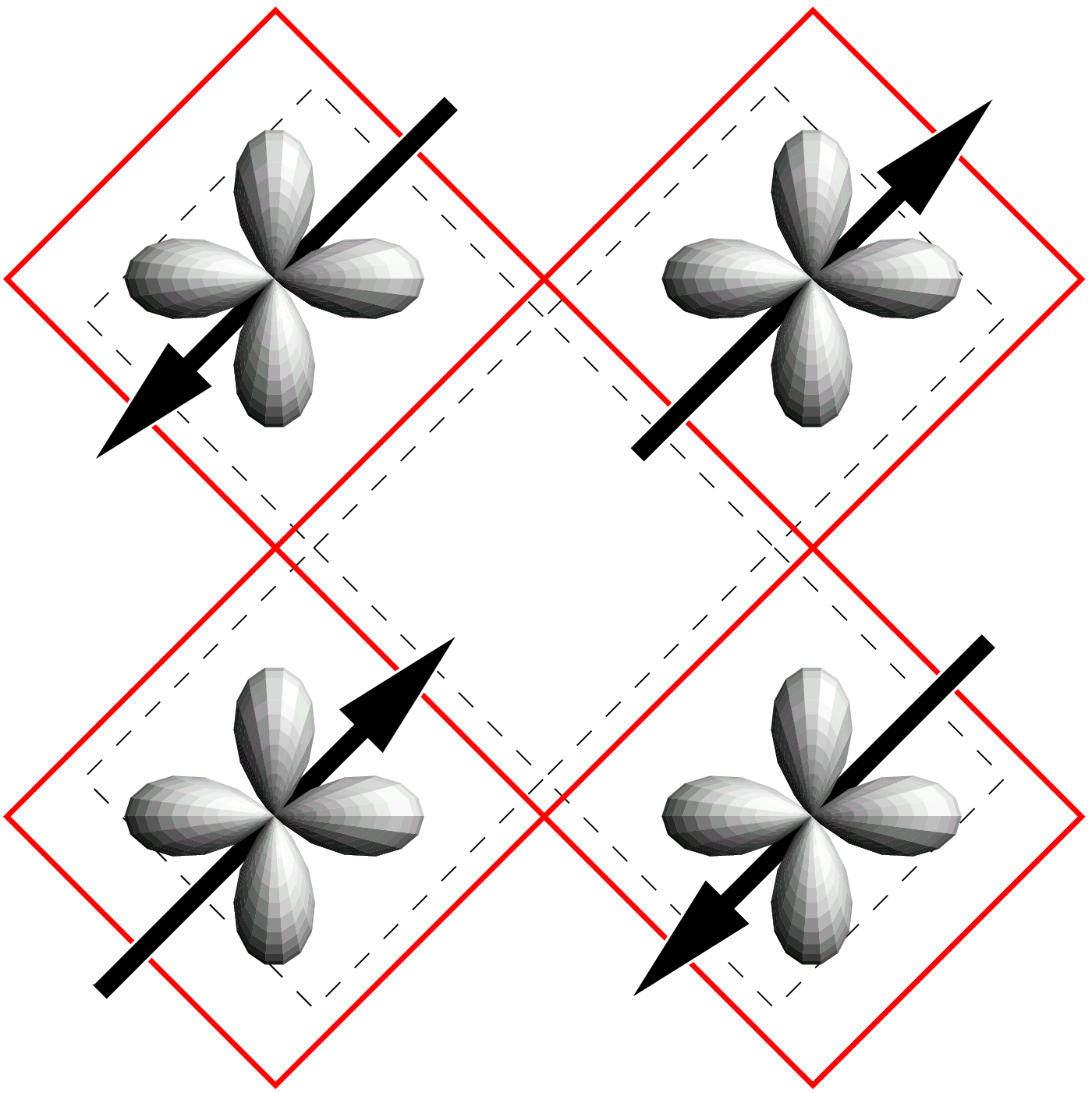} &
      \includegraphics[width=0.2\linewidth]{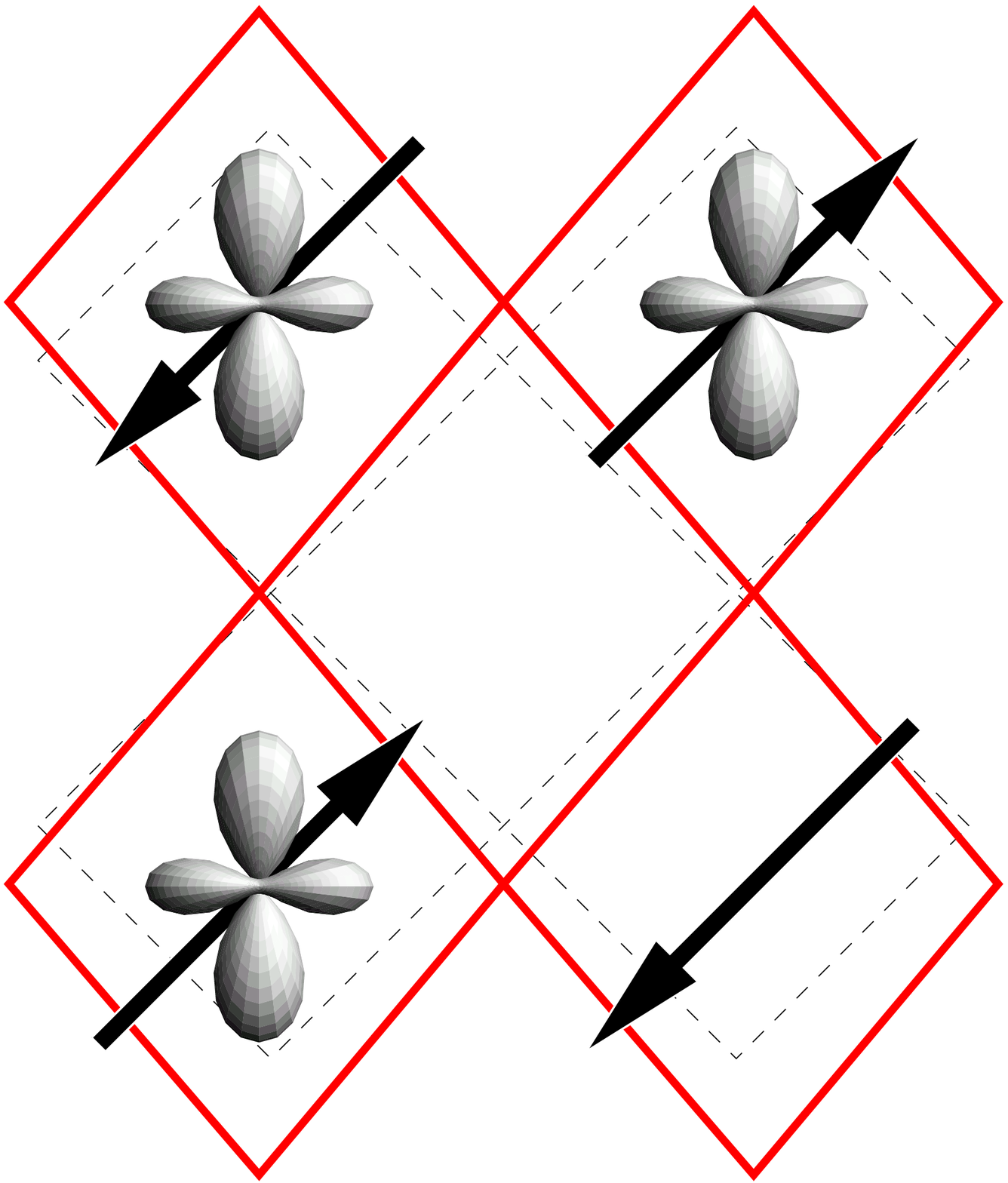} &
      \includegraphics[width=0.2\linewidth]{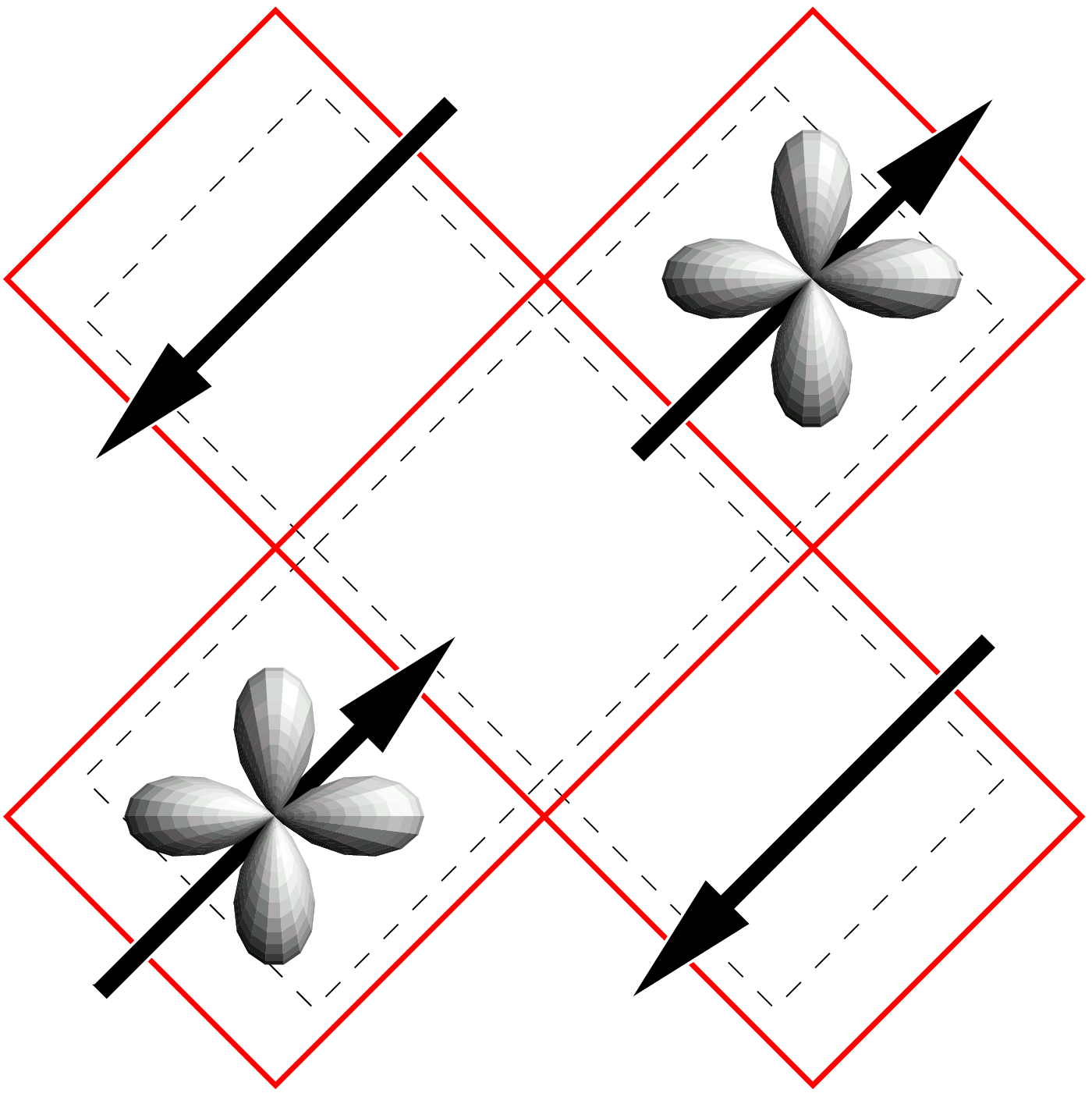}
    \end{tabular}
  \end{center}
  \caption{Schematic summary of the various spin, charge, orbital and 
    lattice patterns calculated within the full model,
    \Eref{fullmicroham}, at weak and strong electron phonon
    interaction and increasing doping.}\label{figorbis}
\end{figure}

Summarising the preceding sections, the complete microscopic model is
given by
\begin{equation}\label{fullmicroham}
  H = \sum_{i,\delta} R_{\delta}(H^{1,z}_{i,i+\delta} + H^{2,z}_{i,i+\delta})
  + H^{\rm EP}\,,
\end{equation}
which contains all essential features of the low-energy physics of the
manganites. The first term accounts for the ferromagnetic
double-exchange interaction, which describes itinerant $e_g$ electrons
that can optimise their kinetic energy by ordering the spin background
formed out of localised $t_{2g}$ electrons. The second term couples
this spin background to the orbital degrees of freedom of the $e_g$
electrons via Heisenberg type exchange interactions modulated by
orbital projectors. It is thus responsible for anti-ferromagnetic
phases, different orbital orderings as well as spin-orbital
correlations. The third term, finally, causes all the polaronic
effects and long range lattice distortions observed in some regions of
the manganite phase diagram, but also affects the orbital ordering.

For analytical methods the above Hamiltonian is far too complex to be
understood in full detail, and even its numerical solution is hard.
Using high performance computers and new optimisation methods for the
lattice degrees of freedom, we were able to study the ground-state
properties of small clusters and to address, in particular, short
range correlations~\cite{WF02l,WF04}. In Figures~\ref{figdope0}
to~\ref{figorbis} we give a short overview of these results. The main
features are the following: In the undoped compounds, e.g.
\dmo{La}{}{}{}, only the second and third term of \Eref{fullmicroham}
are active, and without electron phonon interaction the competition of
the spin-orbital contributions depends sensitively on the values of
the Coulomb and Hunds rule coupling $U$ and $J_h$, respectively. The
dynamics of both subsystems is strongly correlated. With increasing
electron phonon interaction the systems tends to develop static
Jahn-Teller distortions, which also fix the orbital pattern and
subsequently the spin order. Dynamic correlations of spins and
orbitals are suppressed.  Upon doping the ferromagnetic
double-exchange comes into play, and only a substantial polaronic band
narrowing due to strong electron lattice interaction can prevent
ferromagnetic order. The orbital dynamics is coupled mainly to the
charge dynamics, and is less affected by the spin-orbital terms (see
\Fref{figdope1} where also the case $U=\infty$, i.e. the disabling of
most of the spin-orbital terms, is considered). Interestingly, changes
in the spin order are reflected also in the lattice fluctuations, an
effect observed as well in experiment~\cite{BBKLCN98}. Facilitated by
the electron lattice interaction at and above doping $x=0.5$ the
susceptibility of the system to charge ordering dominates the
low-energy behaviour and matching spin interactions lead to
antiferromagnetic order. These features are, of course, well known
from experiment~\cite{CC96,RCMC97,RAKSC02}. The orbital correlations
can show interesting patterns, involving, for instance, complex linear
combinations $|\phi\rangle = \cos(\phi)|\theta\rangle +
\sin(\phi)|\varepsilon\rangle$ of the $e_g$ states combined to
two-site correlations proportional to $|\phi\rangle_i\otimes
|\phi'\rangle_j + |\phi'\rangle_i\otimes |\phi\rangle_j$ with nonzero
imaginary parts of $\phi$ and $\phi'$.  Similar complex correlations
were also found in mean-field type treatments of the
manganites~\cite{Kh00,BK01,MN00}.

Of course, numerical studies of small clusters are of limited value 
if we want to address long-range order or thermodynamic properties.
For the former, mean-field studies of the complete microscopic model
may give some insight and were carried out e.g. for the undoped
compounds~\cite{FO99}. However, since a large number of ordering
patterns usually differs by only tiny amounts of energy, the results
of such calculations often depend on the underlying assumptions. The
study of thermodynamic properties, on the other hand, requires further
simplifications of the model, which will depend on the particular
aspect of the manganite system that we are interested in. Hereafter we
discuss some candidates of such simplified models.

One of the first models that was studied in connexion with the
manganites is the double-exchange Hamiltonian~\cite{Ze51b,AH55,KO72a},
\begin{equation}\label{hamde}
  \fl\quad H^{\rm DE}_{}  = \frac{-t}{2S+1} \sum_{\langle ij\rangle,\sigma}
  \left(a_{i\sigma}^{\dagger} a_{j\sigma}^{}\ c_i^{\dagger} c_j^{} 
    + {\rm H.c.}\right)
  \quad{\rm with}\quad  
  \sum_\sigma a_{i\sigma}^{\dagger} a_{i}^{} 
  = 2 S + c_i^{\dagger}c_i^{} \quad\forall i\,,
\end{equation}
which follows from the first term in \Eref{fullmicroham} by omitting
the orbital degrees of freedom and generalising to arbitrary on-site
spin $S$. Despite its apparently simple structure -- bosons coupled to
spinless fermions -- the model is not solvable directly, but many of
its properties can been studied within the coherent potential
approximation (CPA)~\cite{KO72b,EGK99,GE99} or, even more basic, by
mean-field approximation~\cite{KO72a}. The latter can be obtained by
considering the matrix element for a single bond~\cite{AH55}, $\tilde
t = t (S_T+1/2)/(2S+1)$, and averaging over all orientations of the
total bond spin $S_T$ within an effective mean field $\lambda = \beta
g \mu_B H_{\rm eff}^z$ \cite{KO72a,WLF01a}. The double-exchange model
then reduces to the non-interacting fermion Hamiltonian
\begin{eqnarray}\label{hameffI}
  \fl\qquad  H_{\rm eff}^{\rm DE,1} & = 
  -t\ \gamma_{S}[S\lambda] \sum_{\langle ij\rangle} 
  \left(c_{i}^{\dagger} c_{j}^{} + {\rm H.c.}\right) \qquad{\rm with}\\
  \fl\qquad  \gamma_{S}[S\lambda] & = \frac{S+1}{2S+1} 
  + \frac{\coth((S+1)\lambda)}{2}\left[\coth((S+1/2)\lambda) 
    - \frac{\coth(\lambda/2)}{2S+1}\right]\,,\nonumber{}
\end{eqnarray}
and for a given temperature the parameter $\lambda$ is chosen to
minimise the free energy of the fermions and the spin system.

\begin{figure}
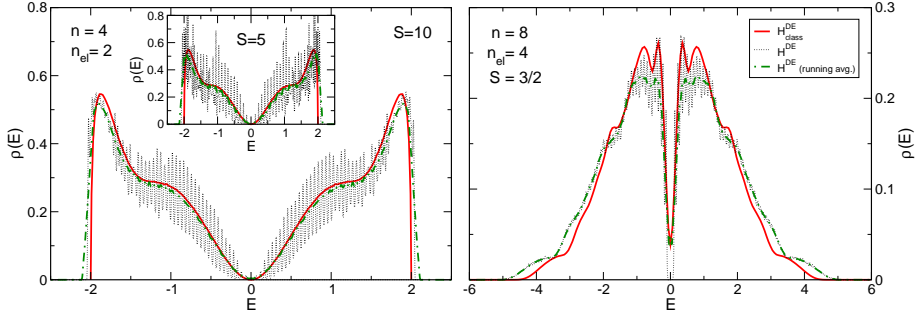

  \begin{center}
    \includegraphics[height=0.2\textheight]{figure17.eps}
    \includegraphics[height=0.2\textheight]{figure18.eps}
  \end{center}
  \caption{Density of non-zero eigenvalues of the quantum double-exchange 
    model with spins of amplitude $S$ (dotted line: discrete spectrum;
    green dot-dashed line: running average) compared to the classical
    double-exchange model (bold red line). Left: Two fermions on a four
    site ring. Right: Four fermions on an eight site
    ring. See~\cite{WLF01a} for more details.}\label{figdxdos}
\end{figure}
If the spin background is less important and the focus is more on the
electrons and their interaction with other degrees of freedom (e.g.
phonons, as discussed below), we may completely neglect the quantum
nature of the spins. The most direct way to find the limit
$S\to\infty$ of \Eref{hamde} is the average over spin coherent
states~\cite{Au94} which, for a classical spin background
parameterised by the polar angles $\{\theta_k,\phi_k\}$,
yields~\cite{KA88,MD96}
\begin{eqnarray}\label{hamdeclass}
  H_{\rm class}^{\rm DE} & = -\sum_{\langle ij\rangle}
  \left(t_{ij}^{} c_{i}^{\dagger} c_{j}^{} + {\rm H.c.}\right)
  \qquad{\rm with}\\
  t_{ij}^{} & = t\left[
    \cos\frac{\theta_i-\theta_j}{2} \cos\frac{\phi_i-\phi_j}{2} 
  + i\,\cos\frac{\theta_i+\theta_j}{2} \sin\frac{\phi_i-\phi_j}{2}
  \right]\,.\nonumber{}
\end{eqnarray}
To access the quality of this approximation, in a recent
work~\cite{WLF01a} we calculated the density of states (DOS) of the
double-exchange model for small finite systems and compared the cases
of quantum and classical spins. As \Fref{figdxdos} illustrates, at
least in a disordered spin background the classical approximation
describes the spectrum very well, even in the case $S=3/2$ that is
relevant for the manganites.

\begin{figure}
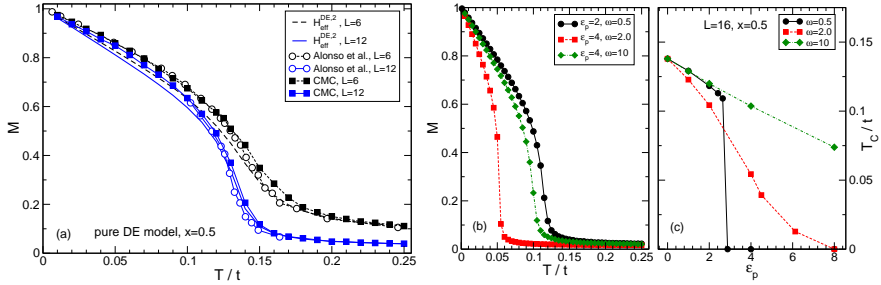

  \begin{center}
    \includegraphics[height=0.18\textheight]{figure19.eps}
    \includegraphics[height=0.18\textheight]{figure20.eps}
  \end{center}
  \caption{(a) Comparison of the magnetisation $M(T)$ of the classical 
    double-exchange model, \Eref{hamdeclass}, and of the effective
    model, \Eref{heffde}, obtained from Monte Carlo
    simulations~\cite{AFGLM01,WFI04p} of $L^3$ site clusters.  (b) and
    (c) Suppression of the ferromagnetic order by electron phonon
    interaction; $\varepsilon_p = g^2/\omega$ denotes the coupling
    strength of the underlying Holstein double-exchange model.
    See~\cite{WFI04p} for more details.}\label{figdmc}
\end{figure}

In analogy to the mean-field model, \Eref{hameffI}, where the spin
background is taken into account only on average, we can also average
over the fermion degrees of freedom to obtain an effective spin
Hamiltonian. For the case of classical spins this leads to
\begin{equation}\label{heffde}
  H_{\rm eff}^{\rm DE,2}  = - J_{\rm eff} \sum_{\langle ij \rangle} 
  \sqrt{1 + \vec S_i\vec S_j}\quad{\rm with}\quad
  J_{\rm eff}  = t\,x (1-x)/\sqrt{2}\,.
\end{equation}
Monte Carlo simulations~\cite{AFGLM01,WFI04p} show that the
magnetisation data and critical temperatures of this model agree
surprisingly well with those for the full classical double-exchange
model [see \Fref{figdmc} (a)].  Summarising these different
approximations we arrive at the conclusion that the double-exchange is
well described by simple effective models, in particular, if we
consider manganites with their almost classical spin amplitude
$S=3/2$.

As discussed above, the double-exchange interaction is responsible for
the ferromagnetic metallic phase of the manganites. However, soon
after the discovery of the colossal magneto-resistance effect
additional electron-lattice interactions were realised to be important
for an understanding of the conductivity~\cite{MLS95,Mi98}.  The third
part of our complete model, $H^{\rm EP}$, contains three different
phonon modes that couple to the orbital degrees of freedom or to the
density, but certainly not all of them are equally important in
different regions of the manganite phase diagram.  For instance,
matching the double-exchange model in \Eref{hamde}, we may again
neglect orbital degrees of freedom and the corresponding phonon modes,
which directly leads to the Holstein double-exchange model
\begin{equation}\label{hamholde}
 \fl\qquad H = \frac{-t}{2S+1} \sum_{\langle ij\rangle,\sigma}
  \left(a_{i\sigma}^{\dagger} a_{j\sigma}^{}\ c_i^{\dagger} c_j^{} 
    + {\rm H.c.}\right)
  +  g\sum_i (b_i^\dagger +b_i^{}) 
  c_i^\dagger c_i^{}
  + \omega\sum_i b_i^\dagger b_i^{}\,.
\end{equation}
Its properties have been studied with a number of methods and seem to
account for many aspects of the metal-insulator transition and of the
transport properties of ferromagnetic-metallic manganites with
$0.25\lesssim x\lesssim 0.5$. A good overview is given e.g. in
Ref.~\cite{Ed02} with a focus on CPA.  Analogous to the pure
double-exchange Hamiltonian, we can also apply a number of further
approximations to \Eref{hamholde}, e.g. classical spins and phonons or
Lang-Firsov type variational treatments of the lattice dynamics. In
\Fref{figdmc} (b) and (c) we illustrate the suppression of
ferromagnetic order by the electron-phonon interaction.  To obtain
these results the spin part of \Eref{hamholde} is approximated by the
effective spin model from \Eref{heffde} and the electron-phonon part
is treated within a modified variational Lang-Firsov
approximation~\cite{LF62r,Fe96}.  The resulting classical model is
simulated with a Cluster Monte Carlo Method~\cite{WFI04p}. Finally
note, that also the polaron problems related to $H^{\rm EP}$ are
interesting by themselves. Neglecting the spin degrees of freedom, our
full Hamiltonian~(\ref{fullmicroham}) reduces to a fermionic hopping
term plus electron phonon interactions of Holstein and $E\otimes e$
Jahn-Teller type. The corresponding few-fermion problems define the
Holstein polaron and $E\otimes e$ Jahn-Teller polaron, whose
properties have been studied numerically~\cite{Ta00,EBKT03}.

Having discussed simplifications mainly of the double-exchange (first)
and polaron (third) part of our microscopic model, we left out the
spin-orbital (second) part so far. However, except for the various
mean-field treatments, a further simplification, which does not alter
the delicate balance of these degrees of freedom, is less obvious. On
the one hand, there are a number of numerical
studies~\cite{MH99,BHMO99,BHO00} which focus on the orbital dynamics
within a fixed spin background. On the other hand, we can study the
interplay of spins and orbitals from a purely theoretical point of
view and consider simplified models like that derived by Kugel and
Khomskii~\cite{KK73r}
\begin{equation}\label{hamkuko}
  H = J\sum_{\langle ij\rangle_\delta}\left[4({\bf S}_i {\bf
      S}_j)(\tau^\delta_i+\frac{1}{2})(\tau^\delta_j+\frac{1}{2}) +
    (\tau^\delta_i-\frac{1}{2})(\tau^\delta_j-\frac{1}{2}) - 1\right]\,,
\end{equation}
where the orbital pseudo spin operators $\tau^\delta$ are related to
our orbital projectors via $R_\delta(P^{\theta/\varepsilon}) = 1/2\pm
\tau^\delta$. Recently, this model was considered in connexion to basic
mechanisms of spin-orbital interaction, e.g. quantum disorder versus
order-out-of-disorder~\cite{KO97,FOZ98,OFZ00}. Note also that there is
a one-dimensional variant of the model, which is exactly
solvable~\cite{Ui70,IQA00}.

\section{Summary}
In this contribution we focussed on a detailed introduction to the
microscopic modelling of colossal magneto-resistance manganites, a
class of materials which is particularly interesting due the close
interplay of charge, spin, orbital and lattice degrees of freedom
characterising its phase diagram and electronic properties.  We
supplemented the derivation by a discussion of various simplified
models and their relation to our Hamiltonian. In addition, we reviewed
some of our previous numerical results for the complete microscopic
model and for its relatives.

\ack We thank D. Ihle and J. Loos for many stimulating discussions and
acknowledge financial support by Deutsche Forschungsgemeinschaft and
Australian Research Council. Calculations were performed at the
facilities of NIC J\"ulich, HLRZ Stuttgart, LRZ M\"unchen, APAC
Canberra and ac3 Sydney.

\section*{References}


\end{document}